\documentclass[]{article}
\usepackage{times}
\usepackage[a4paper, margin=1in]{geometry}

\usepackage{moreverb}
\usepackage{bm}
\usepackage{tikz}
\usetikzlibrary{calc,fit,positioning,arrows,shapes,backgrounds}
\usepackage{natbib}
\usepackage{url}
\usepackage{hyperref}

\newcommand\BibTeX{{\rmfamily B\kern-.05em \textsc{i\kern-.025em b}\kern-.08em
T\kern-.1667em\lower.7ex\hbox{E}\kern-.125emX}}

\title{Bayesian multistate modelling of incomplete chronic disease burden data}

\author{Christopher Jackson {\footnotesize (MRC Biostatistics Unit, University of Cambridge)}\\
Belen Zapata-Diomedi {\footnotesize (Healthy Liveable Cities Lab, Centre for Urban Research, RMIT University, Melbourne)}\\
James Woodcock {\footnotesize (MRC Epidemiology Unit, University of Cambridge)}}
\date{}

\begin{document}

\maketitle

\begin{abstract}
  A widely-used model for determining the long-term health impacts of public health interventions, often called a ``multistate lifetable'', requires estimates of incidence, case fatality, and sometimes also remission rates, for multiple diseases by age and gender.  Generally, direct data on both incidence and case fatality are not available in every disease and setting.  For example, we may know population mortality and prevalence rather than case fatality and incidence.    This paper presents Bayesian continuous-time multistate models for estimating transition rates between disease states based on incomplete data.  This builds on previous methods by using a formal statistical model with transparent data-generating assumptions, while providing accessible software as an R package.   Rates for people of different ages and areas can be related flexibly through splines or hierarchical models.  Previous methods are also extended to allow age-specific trends through calendar time.   The model is used to estimate case fatality for multiple diseases in the city regions of England, based on incidence, prevalence and mortality data from the Global Burden of Disease study.  The estimates can be used to inform health impact models relating to those diseases and areas.  Different assumptions about rates are compared, and we check the influence of different data sources.
\end{abstract}

\section{Introduction}

To inform policies for chronic disease prevention, decision-makers need to quantify the expected impacts of interventions on population health.  This requires knowledge of the current disease burden, e.g. incidence, prevalence, mortality, costs and inequalities, and how these outcomes might change under different scenarios or policies.   Interventions affecting, e.g. diet, physical activity or air pollution exposure may affect multiple diseases in complex ways \citep{mytton2017modelled,briggs2019primetime}.  Disease risks and the impacts of policies may also vary between different geographical areas, e.g. policies relating to transport \citep{iroz2020active}.   Since randomised trials for population health impacts are typically infeasible, theoretical or mechanistic models are required to enable simulation of how outcomes are generated and affected by the interventions \citep{threlfall2015appraisal}.  Different types of models used to simulate health and economic impacts of public health interventions to prevent chronic diseases are reviewed by \citet{briggs2016choosing}.   Many of these require knowledge of age-specific incidence, case fatality and remission rates for each disease of interest, so that each disease can be represented as a three-state transition model (Figure~\ref{fig:states}).   This forms the basis of the ``multistate lifetable'' approach to impact modelling, or the ``proportional multistate lifetable'' \citep{barendregt:pmslt,blakely2020proportional} approach if multiple diseases are modelled independently.  Disease-related outcomes can then be simulated under different policies or scenarios that modify (at least) the disease incidence, either at an aggregate level for a homogenous population with common risks, or through individual-level ``microsimulation'' models that can represent large heterogeneous populations.

\tikzstyle{state} =[draw, rectangle, minimum size = 1cm, text width=2cm, font=\normalsize, align=center, rounded corners, fill=lightgray!5]
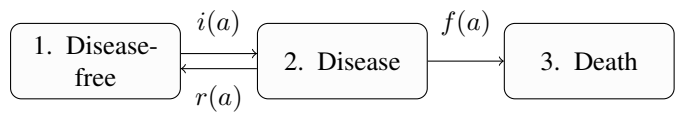
\begin{figure}[htbp]
\begin{tikzpicture}

\node (health) [state] {1. Disease-free};
\node (disease) [state, right = of health] {2. Disease};
\node (death) [state, right = of disease ] {3. Death};

\draw [->] ($(health.east)+(0cm,0.1cm)$) -- ($(disease.west)+(0cm,0.1cm)$) ; 
\draw [->] ($(disease.west)-(0cm,0.1cm)$) -- ($(health.east)-(0cm,0.1cm)$) ; 
\draw [->] (disease) -- (death); 

\coordinate (mid) at ($(health.east)!0.5!(disease.west)$);
\coordinate (mid2) at ($(disease.east)!0.5!(death.west)$);

\node[above=0.2cm of mid] (inc) {$i(a)$};
\node[below=0.2cm of mid] (inc) {$r(a)$};
\node[above=0.2cm of mid2] (inc) {$f(a)$};
\end{tikzpicture}
\caption{\label{fig:states} Multistate disease model with age-dependent incidence, case fatality and remission rates}
\end{figure}

Data to inform those models could come from a range of sources, depending on the location and the detail required on how risks vary between people.  Many diseases, such as cancer, have registries from which population incidence and case fatality can be estimated, however information for subgroups of the population may be restricted.   Commonly, indirect evidence is required.   For example, annual risks of mortality from a disease, for a population including people with and without that disease, are often published by national authorities.  However, case fatality, the risk of death for people with the disease, is less commonly available.   Age-specific prevalence might also be obtained from survey data, and used to infer disease incidence \citep{keiding1991age}.   The Global Burden of Disease (GBD) project \citep{gbd2019} publishes estimates of incidence, prevalence and mortality, but not case fatality, for hundreds of diseases, countries and sub-national regions worldwide. This is a convenient source to inform health impact models that involve many diseases and are designed to be easily adaptable to different geographical settings.  An example is the “Integrated Transport and Health Impact Model" (ITHIM) series of models used to predict the effects of changes in transport behaviours and/policies in different settings \citep{woodcock2014health,de2017health,jaller2020active}. 

Many models to estimate the health impacts of disease prevention interventions \citep[e.g.][]{rehm2009global,cecchini2010tackling,kypridemos2016cardiovascular,mytton2018current}  have used the DisMod II method and software \citep{dismod2}, to estimate the quantities in the model in Figure \ref{fig:states} given a mixture of direct and/or indirect (prevalence and mortality) data. This method estimates the unknown incidence, case fatality and remission rates in Figure \ref{fig:states} through optimising a squared error loss.    In Section~\ref{sec:inference} we clarify the statistical principles behind this method, which have not previously been made explicit.  Essentially, it is a maximum likelihood procedure, however it makes strong assumptions about homogeneity of error variances between ages, and uses some ad-hoc procedures with unclear statistical properties, e.g. to produce rates as smooth functions of age (see Section~\ref{sec:age}).   The assumption that trends through time are common to all ages has been shown to be unsuitable in particular situations \citep{scarborough2016assessing}.  The software is available as a interactive Windows interface, though analysis settings must be specifed with ``point and click'', which makes reproducibility difficult.

An alternative approach to the same estimation problem was described by \citet{flaxman2015}.  This is based on an explicit sampling model for the observed data that represents all structural assumptions, e.g. about how risks are related between different people.   Uncertainties are quantified through Bayesian inference, and the model is fitted by Markov Chain Monte Carlo (MCMC) sampling.   Risks depend on age through linear spline functions, though sensitivity to the choice of the spline knots and smoothness parameter was highlighted.  An empirical Bayes approach was used to share information between areas, by modelling each region separately using a prior obtained from worldwide data.  While Python code to implement models of this form has been published \citep[as DisMod-MR,][]{dismod:mr}, the details of the models implemented in this code, and the quantities that can be estimated from them, are not fully documented, and we are not aware of any applications of the method outside the Global Burden of Disease project.

We develop a modelling framework and software package which combines the statistical formality of the \citet{flaxman2015} methods with the accessibility of the DisMod II tool.  A Bayesian model is used, with two alternative computational methods.  The more accurate, but more expensive, method uses Markov Chain Monte Carlo to sample from the full posterior distribution.  Alternatively, the posterior mode can be determined using optimisation, and supplemented with an approximation to the posterior.  That allows an exact point estimate to be produced, and uncertainty to be quantified, without computational expense.    Our models extend those described by \citet{flaxman2015} in the following ways.   To represent variation in risks with age, instead of a linear spline with pre-specified smoothness, a penalised smooth spline is used that allows the appropriate amount of smoothness to be determined from the data.   To represent variation in risks between areas, as well as empirical Bayes methods, a full hierarchical model can be used which additionally accounts for uncertainty about between-area variations in risk.   The model is also extended to allow risks to depend on calendar time as well as age, while relaxing DisMod II's assumption that the same time trend applies to all ages.  More precise estimates can also be produced in situations where the effect of a predictor (typically gender) can be assumed to be common between areas.   Structural assumptions, such as constant risks under a certain age, can be employed to strengthen inferences from sparse data, and alternative models can be compared against observed data by cross-validation.  The methods are made available in a thoroughly-documented R package at \url{http://chjackson.github.io/disbayes}. 

The Bayesian model, and how it is estimated from data, is fully explained in Section~\ref{sec:models}. The methods we develop are motivated by the demands of health impact modelling in the city regions (or ``combined authority'' areas) of England, described further in Section~\ref{sec:app}.  Published estimates and uncertainty intervals for age-specific incidence, prevalence and mortality for the relevant local areas are obtained from the Global Burden of Disease project.   From these data, our Bayesian model enables case fatality rates to be inferred alongside incidence rates.  This produces a database of estimated disease transition rates that can be used for any multistate lifetable health impact model relating to these diseases and areas.   We compare the plausibility and statistical fit of different model assumptions, and discuss the influence of different sources of data on the results.  Section~\ref{sec:discussion} concludes the paper with a discussion of further challenges of multistate disease modelling for both disease burden estimation and health impact estimation.

\section{Models}
\label{sec:models} 

\subsection{Theoretical disease model} 
\label{sec:model}

We represent a disease as a continuous-time, multistate process with three states: (1) disease-free, (2) disease, (3) death from the disease (Figure~\ref{fig:states}).  
The individual-level disease process is fully defined by the disease incidence $i(a)$, the remission rate $r(a)$ and the case fatality rate $f(a)$, which are assumed to depend on age $a$, defining a continuous-time, age-inhomogeneous Markov model.   For some diseases, remission might be assumed to be implausible, so that $r(a)=0$; an equivalent assumption is that the case fatality from a disease is constant for all times since first onset of the disease until death.  Mortality from other causes is assumed to be independent of disease status, so that the death state for each disease represents only deaths caused by that disease, ignoring other causes, and the rates pertaining to different diseases are estimated independently of each other \citep{barendregt:pmslt}.

Assume further that these rates are constant within integer years of age $a$, so they can be written $i_a$, $f_a$, $r_a$.  For the moment, assume also that the rates do not vary with calendar year, though this assumption is unrealistic for some diseases and will be relaxed later.  Various quantities of interest can be defined as functions of these, as follows.
\begin{itemize}
\item The annual \emph{transition probability matrix} $P_a$, whose $r,s$ entry $P_{a,r,s}$ is the probability that a person is in state $s$ at age $a+1$ given they are in state $r$ at age $a$. $P_a$ is a deterministic function of $i_a$, $f_a$ and $r_a$, which is the solution to a differential equation, given in full in the supplementary Appendix. 

\item The \emph{state occupancy probabilities} $S_a$ or the proportion of individuals in a hypothetical birth cohort (of infinite size) who are in each state at age $a$.  This is a vector with three elements $S_{aj}$, one for each state $j$.  The disease prevalence at age zero is fixed, typically so that everyone is disease-free at age 0.  The state occupancy probabilities at each subsequent age $a+1$ can then be determined by multiplying by the transition probability matrix: 
\[ S_{a+1} = S_a P_a \]

\item The \emph{prevalence} of disease among people who are alive at each age $a$ (in the infinite population):
\begin{equation}
  \label{eq:prev}
  \pi_a = S_{a,2} / (S_{a,1} + S_{a,2})
\end{equation}

\item The population disease-specific \emph{mortality probability} $d_a$ at age $a$, or the probability that a person alive at age $a$ dies from the disease before age $a+1$, which is a function of the disease prevalence $\pi_a$ at age $a$ and the transition probabilities $P_a$ between ages $a$ and $a+1$,
\begin{equation}
  \label{eq:mort}
  d_a = P_{a23} \pi_a + P_{a13} (1 - \pi_a) 
\end{equation}

\end{itemize}

To start with, suppose we are modelling a homogeneous population (e.g. as defined by gender and area) where the rates vary only by age.  In Sections~\ref{sec:inference}--\ref{sec:homog} we discuss methods for estimating unknown quantities in this model from data, and in Section \ref{sec:age} we explain how the dependence on age can be modelled flexibly and efficiently.  In Section \ref{sec:hier} the model will be extended to represent populations from different areas that are related through a hierarchical model, and Section \ref{sec:gender} describes how rates can be related parsimoniously between different subgroups, e.g. by gender.  Section \ref{sec:trends} then explains how trends in risks through calendar time can be modelled.

\subsection{Approaches to inference from data} 
\label{sec:inference}

We assume there are data on mortality from the disease, and at least one of incidence or prevalence, from years of age $a=0,\ldots,A$.  If remission is permitted, there may also be similar data on those assumed to be cured of the disease.   To motivate how we estimate the model in Section~\ref{sec:model} from data, first we explain how the commonly-used DisMod II method \citep{dismod2} works.

In DisMod II, the input disease data are provided as point estimates, e.g. of prevalence or mortality.   The unknown transition rates $i_a,f_a,r_a$ are then estimated by minimising the sum of squared errors between these estimates and those implied by the theoretical model of Section~\ref{sec:model}.   The statistical theory behind this procedure is not made explicit, but it is equivalent to maximum likelihood estimation, under the assumption that the point estimates are observations from a normal sampling distribution with mean defined by the true quantity, and a variance that is the same for each age.  In reality, however, the error variance will be greater for ages at which the input quantities are more uncertain due to being obtained from smaller samples (e.g. because the population at risk varies with age).

Uncertainty in DisMod II can be quantified by supplying distributions around the inputs.  Monte Carlo simulation from these distributions is then used to determine the implied distribution of the estimated transition rates.  Therefore, if the user supplies the correct sampling distributions behind their input data, this is equivalent to a bootstrap method to estimate the sampling distribution of the transition rate estimators.   No guidance is given, however, on how a user can derive those input distributions to reflect the knowledge that is available in real situations.  As well as knowledge about sampling variability, this includes ``structural'' knowledge about how rates for different people (e.g. of different ages, genders and areas) are expected to be related to each other, and expected biases or inconsistencies between data sources.   In DisMod II, structural knowledge can be introduced by ad-hoc methods, e.g. by smoothing age-specific point estimates after estimation to reflect that rates are expected to be similar between ages, though, as discussed in Section~\ref{sec:age}, this does not use all information efficiently.

Therefore, instead of expressing the problem as minimising a loss, we prefer to define a statistical model from which we assume the data are generated, and where all data-generating assumptions are made explicit.  This produces a likelihood function for the unknown parameters, which can either be maximised or combined with a prior in Bayesian inference.  Uncertainties can then be automatically represented, either through obtaining the covariance matrix of the maximum likelihood estimates, or in the posterior distribution in the Bayesian approach.   As in \citet{flaxman2015}, we use Bayesian inference, and any relations between rates from different people are explicitly described in the model or as prior distributions.   The details of this model, and how we extend on the \citet{flaxman2015} method, are described in the following sections.

\subsection{Binomial likelihood for a homogeneous population} 
\label{sec:homog} 

Instead of point estimates, the data are expressed as counts, which explicitly acknowledges that they were obtained from observing disease outcomes from a finite population.  Specifically, we suppose we have numerators and denominators, as follows.
\begin{itemize}
\item[incidence:] given a population at risk, of size $n_a^{(inc)}$, $y_a^{(inc)}$ of these are observed to get the disease within the next year,
   
\item[mortality:] given a population at risk, of size $n_a^{(mort)}$ (with or without the disease), $y_a^{(mort)}$ of these are observed to die from the disease within the next year,
  
\item[prevalence:] from a sample of $n_a^{(prev)}$ individuals, $y_a^{(prev)}$ are known to have the disease, and $n_a^{(prev)} - y_a^{(prev)}$ are known to not have the disease, at age $a$. 
\end{itemize}
Remission data $y_a^{(rem)},n_a^{(rem)}$ may also be available in a similar form, if remission is permitted.

As shown in Appendix Section B, if the data are published as point estimates, with a (published or assumed) measure of uncertainty around the estimates, we can derive approximately equivalent numerators and denominators of the required form.    Furthermore, the data may be published as estimates for coarser age groups (5 or 10 year bands) while estimates are required for one year of age.  A procedure for smoothly disaggregating counts for age groups into estimated year-specific counts, with no loss of information, is discussed in Appendix Section C. 

Under the theoretical disease model, if individuals in the population are assumed to be independent with identical risks, these data are generated as 
\begin{itemize}
\item[incidence:] $y_a^{(inc)} \sim  Binomial(n_a, 1 - P_{a11})$, where $1 - P_{a11}$ is the probability of getting the disease at some time within a year, which we will call the \emph{incidence probability}.
\item[mortality:]  $y_a^{(mort)} \sim  Binomial(n_a, d_a)$, where the disease-specific mortality $d_a$ is a deterministic function of the incidences and case fatalities $\{i_j,f_j\}$ for ages $j$ up to $a$ (Equation \ref{eq:mort}).
\item[prevalence:]  $y_a^{(prev)} \sim Binomial(n_a^{(prev)}, \pi_a)$, where $\pi_a$ is the theoretical prevalence, defined as a deterministic function of the incidences and case fatalities~(Equation \ref{eq:prev}).
\item[remission:]  $y_a^{(rem)} \sim Binomial(n_a^{(rem)}, P_{a,2,1})$, where $P_{a,2,1}$ is the annual transition probability from disease to health. 
\end{itemize}

The Binomial model above defines a full likelihood $L(\bm\theta | \mathbf{y})$ for the data $\mathbf{y} = \{y_a^{(mort)}, y_a^{(prev)},y_a^{(inc)},y_a^{(rem)}\}: a=0, \ldots, A$ and parameters given by the rates $\bm\theta = \{ i_a, f_a, y_a \}: a=0,\ldots,A$.

To estimate the rates, in theory, this might be maximised as a function of the $\bm{\theta}$, independently for each population that we want to describe (e.g. by area and gender).  Similarly, Bayesian inference might be used with independent priors for the rates at each year of age $a$.   However, this ignores the knowledge that risks at adjacent ages will be similar, and the data will often be too weak to allow all of the age-specific parameters to be identified solely from age-specific data --- in particular, the case fatality rates $f_a$ will be too weakly informed by the data on mortality, in settings and ages where the disease is uncommon.

Instead, we build in structural assumptions that will define how rates for particular people are assumed to be similar to each other.  Firstly, Section~\ref{sec:age} defines a model for how rates depend on age.  Later we will define models that can be used to describe how rates vary between different contexts such as geographical areas (Section \ref{sec:hier}) or by gender (Section \ref{sec:gender}).    Building in plausible structural relations between parts of the data in this way improves identifiability of the estimates and increases their precision.   Rates may also vary through calendar time (Section~\ref{sec:trends}).   A further challenge is that the datasets informing incidence, mortality, prevalence and remission might be obtained from different sources, hence describe populations with slightly different epidemiology (Section~\ref{sec:bias}).

\subsection{Spline models for dependence on age}
\label{sec:age} 

Since the rates for similar ages are expected to be similar, the case fatality and incidence are assumed to be smooth, nonlinear functions of age.   The model for case fatality is:
\begin{equation}
  \label{eq:age}
  \log(f_a) = \sum_{k=1}^K \beta_k g_k(a)  
\end{equation}
where $g_k()$ are spline basis functions.  The models for incidence and remission rates have an identical form, but governed by different parameters.   

Any spline basis can be used in theory. \citet{flaxman2015} describe linear splines with pre-specified knots and smoothness penalties.  We extend this by allowing the appropriate shape and smoothness to be estimated from the data as part of the model fit, and we use a more flexible ``thin plate'' spline, parameterised as in \citet{jagam} to facilitate Bayesian modelling.   The exact form of this basis is complex, and described in full in \cite{wood2017generalized}.  A large number $K=10$ of basis terms are included to ensure high flexibility if required.  Note this does not necessarily give an over-parameterised model, because the appropriate level of flexibility is estimated, as we now describe.   The first two terms represent an intercept and slope, $g_1(a) = 1, g_2(a) = a$.  The remaining terms represent departures from a linear relationship of the log rate with age, and their coefficients are assumed to be exchangeable draws from a common distribution $\beta_k \sim N(0,\lambda_0^2)$ for $k\geq 3$.   The prior standard deviation $\lambda_0$ controls the degree of smoothness, with the curve tending towards a straight line for $\lambda_0 \rightarrow 0$ and becoming more flexible for large $\lambda_0$.   A prior is placed on $\lambda_0$ (we use a Gamma(2,1) to facilitate estimation of the posterior mode, see \citet{gelman:bda3}, Chapter 10) hence Bayesian updating of this parameter allows the appropriate amount of smoothness to be estimated from the data. 

If we are modelling a homogeneous population (e.g. a specific area and gender) we can then complete the model by defining priors for the intercept $\beta_1$ and slope $\beta_2$.  In the application, we use vague $N(0,100^2)$ priors. 

The parameters $\beta_k$ and $\lambda_0$ defining the age curve are included in the likelihood and estimated as part of the Bayesian model.  This differs from Dismod II's approach to smoothing, where age-specific rates are estimated independently, and the resulting estimates can then be smoothed.  While that ensures that estimates from adjacent ages are similar, an advantage of our approach, where the rates and the functional form relating them are estimated in a single step, is that it allows the improvements in precision from borrowing information between ages to be accounted for in reduced uncertainty around the estimates.

Two further assumptions about age dependence of disease rates are used, depending on the disease.    Firstly, we define an age $a_{base}$ below which rates are assumed to be constant.   For example, incidence may be zero or low for some diseases at younger ages.  Or relatedly, if the prevalence of the disease is low at younger ages, then there will be no information about case fatality at these ages, meaning that the model needs to constrain how case fatality depends on age.    Secondly, for some diseases, we impose the additional constraint that the rates are \emph{increasing} with age (after the period when rates are constant).  This is enabled by using the same form of spline model, but for the log \emph{increments} in rates with each year of age, rather than the log rates. 

An alternative model with independent vague priors for each year of age was also found to be useful for model development --- since if the age-specific posterior for case fatality is identical to the age-specific prior in that model, we can deduce that the data provide no information about that particular age, confirming that additional structural or prior assumptions would be needed instead.


\subsection{Hierarchical model for variations between areas} 
\label{sec:hier} 

For some diseases and areas, the information in a single area alone may be too weak to give sufficiently precise estimates of case fatality or incidence.    However, a single area's data can be strengthened through the information provided by other areas.  As an alternative to aggregating the data over areas, a hierarchical model can be used, which ``partially pools'' weak data from one area with the average from other areas.    In DisMod II there is no hierarchical modelling ability. \citet{flaxman2015} implemented partial pooling by using estimates from data aggregated over multiple areas to define a prior for the area-specific rate in models fitted to area-specific datasets independently.   They also investigated a full Bayesian hierarchical model, which represents between-area variations as random effects and has the advantage of fully accounting for uncertainties about between-area variations, but found this model to be computationally prohibitive.   

Here we describe a similar fully Bayesian hierarchical model, and implement it using Hamiltonian Monte Carlo methods (see Section~\ref{sec:comp}), which are designed to explore correlated posterior distributions more efficiently than the Metropolis-Hastings methods that were used by \citet{flaxman2015}.   In our applications, convergence, and 1000 uncorrelated samples from the posterior, were achieved in 1-5 hours of running time, depending on the structural assumptions.  We also implement an efficient approximation to the full hierarchical model (Section~\ref{sec:comp}) that ran within minutes in our examples.

In this setting, there are data $\mathbf{y} = \{y_{i,a}^{(mort)}, y_{i,a}^{(prev)},y_{i,a}^{(inc)}\}$ published by area $i$ as well as age $a$, and are related to area-specific rates $\theta = \{ i_{i,a}, f_{i,a}, y_{i,a} \}$ through the model described in Sections~\ref{sec:model}--\ref{sec:homog}, leading to a joint likelihood $L(\theta | \mathbf{y})$ over ages and areas.   As before, the rates are smooth spline functions of age, and the models for incidence and case fatality have the same structure, but with different parameters for each area.   For case fatality, $\log(f_{i,a}) = \sum_{k=1}^K \beta_{i,k} g_k(a)$,  and for incidence, $\log(i_{i,a}) = \sum_{k=1}^K \beta_{i,k}^{(inc)} g_k(a)$.  Remission rates might be treated in the same way in principle, but given the available data in the application, these are assumed to be constant over areas as well as ages.    Then to make this model hierarchical, the intercept term $\beta_1$ in the spline function becomes a \emph{random effect} $\beta_{i,1}$, and is given a distribution that represents the variation between areas in the level of risk.  In the application in Section~\ref{sec:app},  the random intercepts for log case fatality are assumed to be exchangeable, $\beta_{i,1} \sim N(b_1, \lambda_1^2)$.  A random slope could be defined similarly by placing a distribution on $\beta_{i,2}$, however in the application this was judged to be unnecessary, and a common slope $\beta_{i,2}=b_2$ was used.

The sharing of information can be controlled by the prior distribution for the between-area random effect standard deviation $\lambda_1$.  To allow this prior to be defined transparently, we imagine ``high-risk'' and ``low-risk'' areas, defined by the 2.5\% and 97.5\% quantile of the distribution of case fatality, and place a prior guess on the ratio in case fatality between a high and low risk area, and a plausible upper limit on this ratio.  A guess of a 5-fold ratio, with an upper limit of a 50-fold ratio, is used later in the application.  A gamma prior distribution for $\lambda_1$ is then obtained by a numerical search for the gamma shape and scale parameters that correspond to this prior mean and 97.5\% upper quantile.    The nonlinear terms $\beta_{i,3},\ldots,\beta_{i,K}$ are given identical $N(0, \lambda_0^2)$ priors, where $\lambda_0$ again controls the degree of smoothness, which is assumed to be the same for all areas $i$.  A vague $N(0,10^2)$ prior is used for the mean intercept, $b_1$, and a $N(5,5^2)$ for the common slope $b_{2}$ --- more informative than the priors for $\beta_1,\beta_2$  used in~\ref{sec:age}, to facilitate computation, though still covering an extremely wide range.

\subsection{HIerarchical models with additive gender and area effects}
\label{sec:gender}

Typically we will want to estimate how disease risks differ by gender as well as age, and there will be data by age $a$, area $i$ and gender $j$, e.g. outcome counts $\mathbf{y} = \{y_{i,a.j}^{(mort)}, y_{i,a.j}^{(prev)}, y_{i,a.j}^{(inc)}\}: a=1, \ldots A; j=1,2$ and corresponding denominators.  The data for a specific $i,j$ are again assumed to be generated from the same theoretical disease model, with parameters given by the rates $\theta = \{ i_{i,a.j}, f_{i,a,j}, y_{i,a,j} \}: a=1,\ldots,A$.    Instead of analysing the datasets for each gender independently under the previously-described models, more precise estimates might be produced under an assumption that the effect of gender is \emph{independent of the effect of the area}.    Then, after adding a third index $j$ in the model for case fatality: $\log(f_{i,a,j}) = \sum_{k=1}^K \beta_{i,j,k} g_k(a)$,  $\log(i_{i,a,j}) = \sum_{k=1}^K \beta_{i,j,k} g_k(a)$, we would have $\beta_{i,j,k} = \beta^{(area)}_{i,k}$  for females ($j=1$), and $\beta_{i,j,k} = \beta^{(area)}_{i,k} + \beta^{(male)}_{k}$ for males $(j=2)$.

$\beta^{(area)}_{i,k}$ are assigned the same priors as the $\beta_{i,k}$ in Section~\ref{sec:hier}.  In our case study, a normal prior with mean 0 and standard deviation 0.82 is used for $\beta^{(male)}_{1}$ and $\beta^{(male)}_{2}$, the gender effect on intercepts and slopes of the log-linear model, representing a 95\% prior probability that the female/male rate ratio, and the female/male ratio in age trends, are between 0.2 and 5.  For $k>2$, the effects $\beta^{(male)}_k$ of gender on each of the $k$th spline basis coefficients are given $N(0, {\lambda_0^{(male)}}^2)$ priors, enabling the gender effect to deviate flexibly from a linear function of age. $\lambda_0^{(male)}$ determines the flexibility of this function, and can be fixed or given a prior and estimated.

\subsection{Time trends}
\label{sec:trends}
The model presented so far, specifically the recursive definition in Section~\ref{sec:model} of the state occupancy probabilities $S_a$ and prevalence at age $a$ in terms of risks at all previous ages, assumes that risks for a particular age $a$ do not change through calendar time.      Therefore, for example, when using data for a mixed population from one year (2017 say), the model assumes that the risk faced by a person of someone of age 50 in 2017 is the same as the risk faced in 1997 by someone aged 70 in 2017 (who were 50 in 1997).   This is unrealistic for many diseases and populations.  From data by age group for one calendar year, it is not possible to distinguish trends through time from differences between age groups.   In theory, these might be distinguished by extending the framework in Section~\ref{sec:homog}  to include cross-sectional data from multiple years, however, we expect that identifiability and sampling from the posterior distribution would be difficult.

Therefore, instead of using our model to \emph{estimate} time trends from data, we extend the model to include published point estimates of time trends as constants, to determinine the \emph{consequences} of these trends for our inference of current incidence and case fatality from current data.    The rates $i_a,f_a,r_a$ now describe risks in the current year, that is the year represented by the data in Section~\ref{sec:homog}.   Risks in previous years are defined by multiplying the current risks by a constant ratio determined from literature on disease trends.     This allows the model to be extended so that incidence and case fatality depend on both age and (calendar) year, $i_{a,y}=\rho^{(i)}_{a,y} i_{a}$, $f_{a,y}=\rho^{(f)}_{a,y}f_{a}$, where the $\rho$ indicate the ratios determined from literature, assumed to be known perfectly, and the year of the data is labelled $y=100$, thus the earliest year represented is 100 years prior to this, $y=0$, the year of birth for people 100 years of age in the current data.   The models in DisMod II allowed time trends in a similar way, but the risk ratios $\rho_{a,y}$ were restricted to be the same for all ages $a$, while our model allows them to be age-dependent.  The models in \citet{flaxman2015} assumed no time trends.

The age and year-dependent rates can then be used to define age and year-dependent transition probabilites and state occupancy probabilities $P_{a,y}$, $S_{a,y}$ that are assumed to generate the observed data.   The data are the same as before --- note that they are only available for year $y=100$.  The mortality data, for example, follows the same binomial model defined in Section~\ref{sec:homog}.  The probability $d_a$ governing this model, however, is now defined in terms of age and year-specific transition probabilities $P_{a,y,r,s}$, and the prevalence $\pi_{a,y} = S_{a,y,2} / (S_{a,y,1} + S_{a,y,2})$ at $y=100$.   For each age $a$,
\[
  d_a = P_{a,y=100,2,3} \pi_{a,y=100} + P_{a,y=100,1,3} (1 - \pi_{a,y=100})
\]
$P_{a,y,r,s}$ is the probability that a person is in state $s$ at age $a+1$ and year $y+1$ given they are in state $r$ at age $a$ and year $y$. The matrix $P_{a,y}$ is defined as a function of $i_{a,y}$ and $f_{a,y}$, the same function used to define the year-independent matrix.   $S_{a,y}$ is the vector of probabilities that a person born in year $y-a$ (thus is of age $a$ at year $y$) occupies each of the states at year $y$, again defined recursively as $S_{a,y} = S_{a-1,y-1} P_{a-1,y-1}$, where $S_{0,y}$ is fixed, e.g so that people are disease-free at age 0 with probability 1, and we need to know $P_{b,y}$ for each pair of $(b,y) = (a-1,99),(a-2,98),\ldots,(0,100-a)$, for each of $a=1,\ldots,100$.

This approach is illustrated in Section~\ref{sec:app} for ischemic heart disease.

\subsection{Model checking and comparison}
\label{sec:bias}

The model in section \ref{sec:homog} assumes that the disease rates that generate the three or four different data sources (incidence, prevalence, mortality and remission)  are the same.   In reality, the data on these sources may have been obtained from different populations, perhaps at different times or places, with different disease epidemiology.  Using the model in that case will produce ``average'' parameter estimates which describe an unclearly-defined ``mixed'' population.  Ideally, the inferences from the model should have a transparent connection to a dataset that describes a known population. 

Consistency between data sources can be checked by comparing observed data with corresponding estimates obtained from the model.  For example, suppose there is direct data on incidence, and the model is used to jointly estimate incidence and fatality rates from mortality, prevalence and incidence data.  We could compare observed incidence data with incidence estimates from the model, and if they disagree to an extent that is unexplainable by sampling variation, then this suggests that the prevalence and mortality data provide evidence about incidence that conflicts with the direct data on incidence.  Conflicts might occur if time trends in incidence have not been properly accounted for, so that past incidence (which generated the current prevalence and mortality data) is different from current incidence, or if the underlying populations behind the data sources are otherwise different. 

Biases might be handled by better adjustment for time trends, by excluding the biased data source if it is not necessary, or by adding model parameters that describe the differences between the data sources (akin to the ``node splitting'' ideas of \citet{presanis2013conflict}).  However there is a limit to how much we can accommodate biases without making it impossible to learn from the data, e.g. information about both mortality and prevalence is required to be able to estimate case fatality. 

To compare different models for the same data, e.g. a simpler or more constrained model (that might be biased) and a more complex or flexible model (that may be estimated imprecisely), their predictive ability can be assessed using leave-one-out cross-validation, via the method and R package of~\citet{loocv,loor}.  For each observation $i$, this method estimates $elpd_i$, the expected log predictive density (ELPD), a measure of the accuracy with which a model would predict the $i$th observation if it were left out when fitting the model.  In our application, the $i$th observation is defined as the outcome count $y_a^{(out)}$ (Section~\ref{sec:homog}) for the $i$th combination of age and outcome type $out$ (incidence, mortality, prevalence or remission), in addition to area or gender if modelled.  The sum $\sum_i elpd_i$, over all $i$ for some outcome, gives an overall measure of the accuracy with which a model predicts that outcome. 

\subsection{Computation and software} 
\label{sec:comp}

The joint likelihoods and priors are fully defined by the data model from Section~\ref{sec:homog} together with the age dependence model~(Section \ref{eq:age}), and potentially also the models to combine data from different areas (Section~\ref{sec:hier}) or genders (Section~\ref{sec:gender}), or to account for time trends (Section~\ref{sec:trends}).  In each case, the posterior requires simulation to describe, which is done here using two alternative methods.   The more expensive but more accurate approach is to sample from the exact posterior by a Markov Chain Monte Carlo algorithm, similarly to \citet{flaxman2015}.   A faster alternative is to use numerical optimisation to find the mode of the posterior.  When the mode is found, a sample can be produced instantly from a multivariate normal approximation to the posterior, defined through a second-order approximation to the log density function on the unconstrained parameter space (see~\citet{gelman:bda3}, Chapter 4).   In these models, the normal approximation can be computed practically instantly, at around 1/100 of the running time of the equivalent MCMC procedure.   The optimisation method is just as fast as the optimisation used by DisMod II, with the advantage of being based on a explicit statistical model, as described in~\ref{sec:inference}.  A further advantage is that approximate uncertainty intervals are produced more efficiently than in Dismod II's procedure of repeatedly computing estimated rates for each resampled input dataset.

An R package \texttt{disbayes} was developed, which implements any of the models described here, using the Hamiltonian MCMC and optimisation procedures available in the Stan software, used through its \texttt{rstan} interface \citep{rstan}.  This package is available from \url{http://chjackson.github.io/disbayes}, with thorough documentation and worked examples, and is described in more detail in Appendix Section D.


\section{Application: estimating multistate lifetable data for city regions in England}
\label{sec:app}

We aim to estimate disease transition rates that can inform health impact models for use in the city regions of England.  These include models for changes in active transport, that build on the “Integrated Transport and Health Impact Model" series of models \citep[from][]{woodcock2014health,de2017health,jaller2020active}.  Multi-state lifetable models are required to simulate long-term population disease outcomes under scenarios where incidence and case fatality are modified by changes in risk, e.g. due to increased physical activity.   This requires area-specific estimates of incidence and case fatality of multiple chronic diseases that may be affected by changes in the relevant risk factors.    While direct data on incidence is available for the required diseases and areas, case fatality is not.   In this section, we show how our model is used to infer case fatality, along with incidence, from data on prevalence, incidence and mortality and remission.   Full code and data to reproduce the analyses in this section is available at \url{http://github.com/chjackson/disbayes}, in the \texttt{metahit} folder.

\subsection{Disease data} 

Data on incidence, prevalence, and mortality for chronic diseases in the year 2019 are obtained from the Global Burden of Disease (GBD) project \citep{gbd2019}  (URL: \url{http://ghdx.healthdata.org/gbd-2019}) by gender, 5-year age groups (from 0-4 to 95-99) and local authority districts within 9 English city regions.  The chronic diseases included are those with evidence of an association with physical activity, air pollution or noise exposure.  For concise presentation, in this paper we illustrate the analyses for the six of these that are associated with the greatest numbers of deaths (ischemic heart disease, stroke, chronic obstructive pulmonary disease, lung cancer, colorectal cancer, breast cancer and dementia), and two less common diseases (stomach and uterine cancer).   Seven additional diseases are included in the full analysis available online.

The published data consist of estimated probabilities with 95\% credible intervals.  These were converted to an approximate numerator and denominator, using the procedure described in Appendix Section B.   Those counts, for 5-year age ranges, are then disaggregated to give smoothly-varying 1-year counts, using the methods implemented by~\citet{RJ-2013-028}. This results in one-year counts that are constrained to sum to the five-year totals, but vary smoothly by age.  These counts are then aggregated over local authority districts within each city region (as defined in \url{https://www.ons.gov.uk/economy/economicoutputandproductivity/output/articles/cityregionsarticle/2015-07-24#appendix}), to produce counts for a city region.

Information about remission rates for each of the cancers is obtained from published estimates and confidence intervals for 10-year survival probabilities for England, published by age ranges (of width 10 years or more)~\citep{ons:cancer:survival}.  As in the GBD study \citep{gbd2019}, we assume that 10-year survival implies that remission happened within those 10 years.  An annual remission probability $r$ can then be deduced from the 10-year survival probability $1 - (1-r)^{10}$. The probabilities are converted to numerators and denominators, disaggregated to years of age, and assumed to be equal for all city regions.

\begin{figure}
  \includegraphics[width=0.9\textwidth]{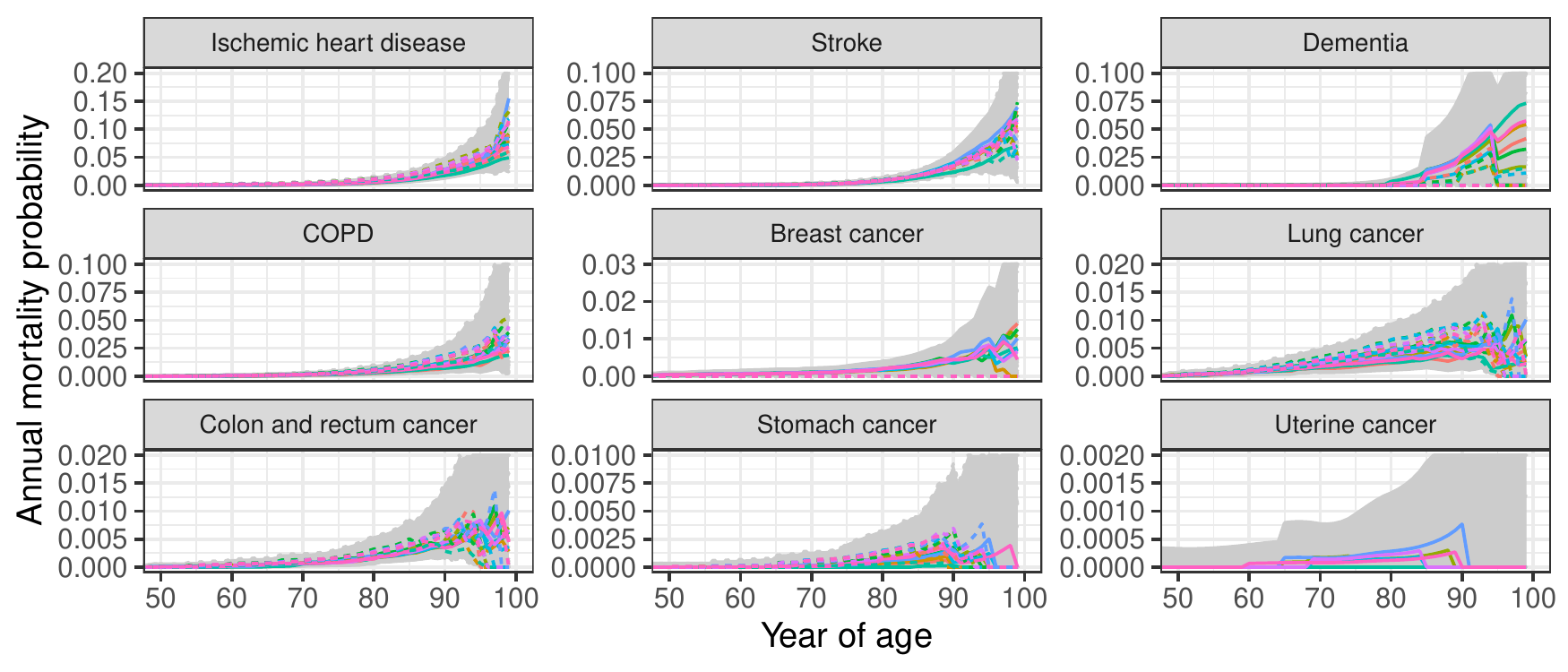}
  \includegraphics[width=0.9\textwidth]{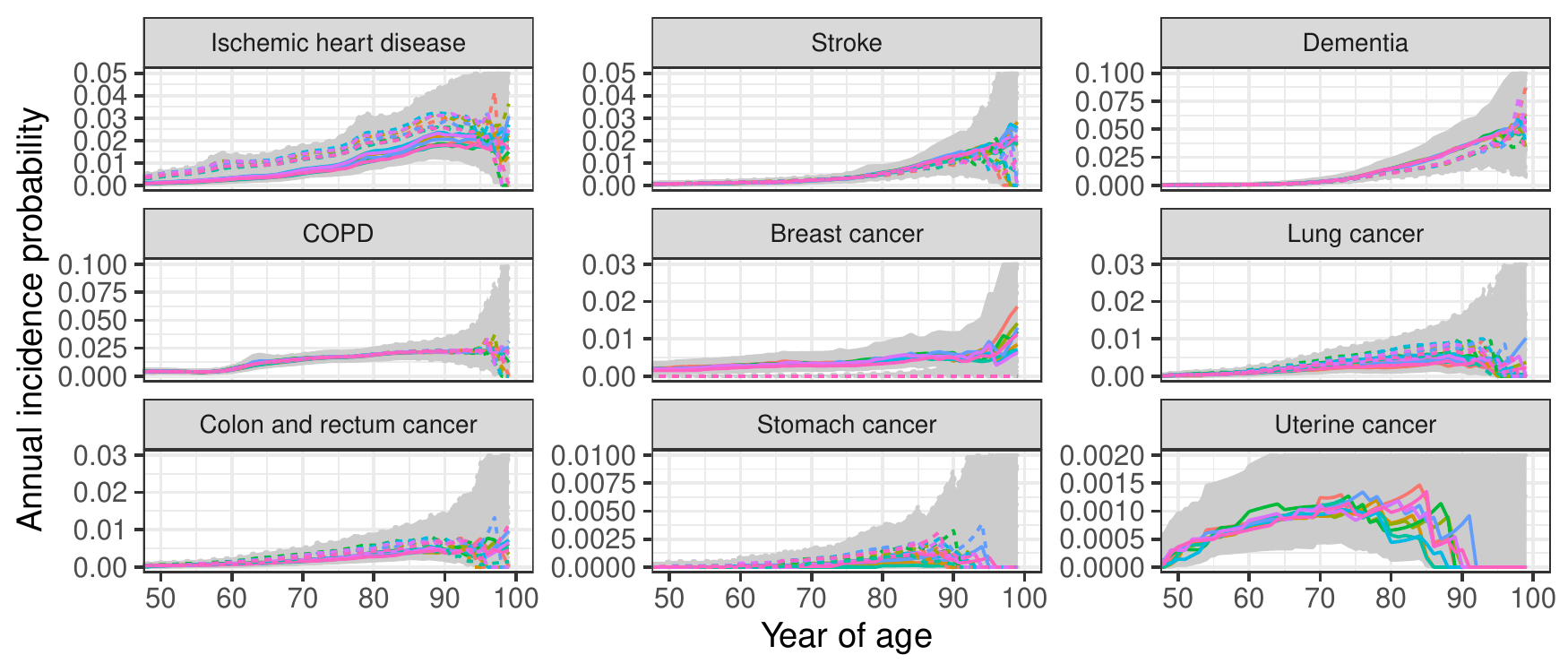}
  \includegraphics[height=2.5in]{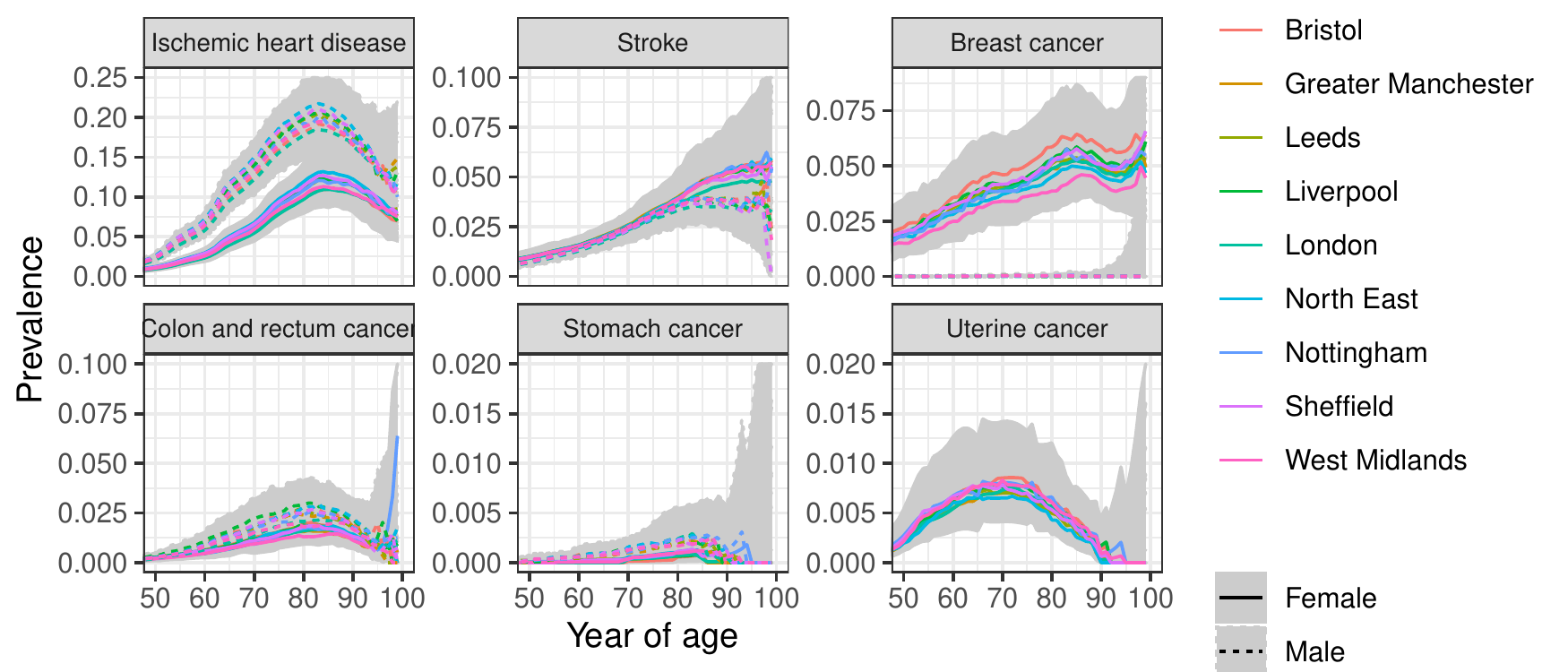}
  \caption{\label{fig:data}~Mortality, incidence and prevalence by disease, age, gender, and city region in England, 2019. Estimates from the Global Burden of Disease, smoothly disaggregated from 5-year age groups.  Shaded areas encompass the published 95\% credible intervals from all city regions (some are truncated above). }  
\end{figure}

Figure~\ref{fig:data} illustrates these data, in the form of estimated probabilities and 95\% credible intervals that are approximately equivalent to the numerators and denominators. This shows how the mortality, incidence and prevalence for each disease changes with age, for people over 50, compared between city regions and genders.   Both the incidence and mortality are lowest for uterine cancer and stomach cancer.   Differences between city regions are moderate compared to differences between ages.  The extent of uncertainty tends to increase with age, and most estimates are highly uncertain beyond age 90.

\subsection{Estimates from models fitted to areas independently} 
\label{sec:app:homog}

For each disease, city region and gender, the model of Section~\ref{sec:homog} was fitted to estimate the case fatality given the mortality, incidence and prevalence.   Case fatality, incidence and remission (where included) were smoothed as functions of age~(as in Section~\ref{sec:age}), while remission rates for diseases other than cancer were assumed to be zero.  The baseline age $a_{base}$, below which rates were assumed constant, was varied according to the disease and the data that were available for younger ages, with $a_{base}=70$ for dementia, uterine cancer and stomach cancer, $30$ for ischemic heart disease and $50$ for all other diseases. Case fatality rates were also constrained to be increasing with age for dementia, stomach cancer, lung cancer and uterine cancer.  For uterine cancer and stomach cancer, city-region-specific estimates of case fatality could not be obtained due to the small numbers of deaths per area and age (75 deaths overall for uterine cancer, and less than around 10 cases of stomach cancer per year of age and city region) leading to difficulties with estimation.   Instead, only national estimates were produced for these diseases.   For a small subset of diseases and areas, the parameters $\lambda_0$, determining the smoothness of the age-dependence, were fixed at their posterior modes (obtained from optimisation) to stabilise MCMC estimation. 

\begin{figure}
  \includegraphics[width=\textwidth]{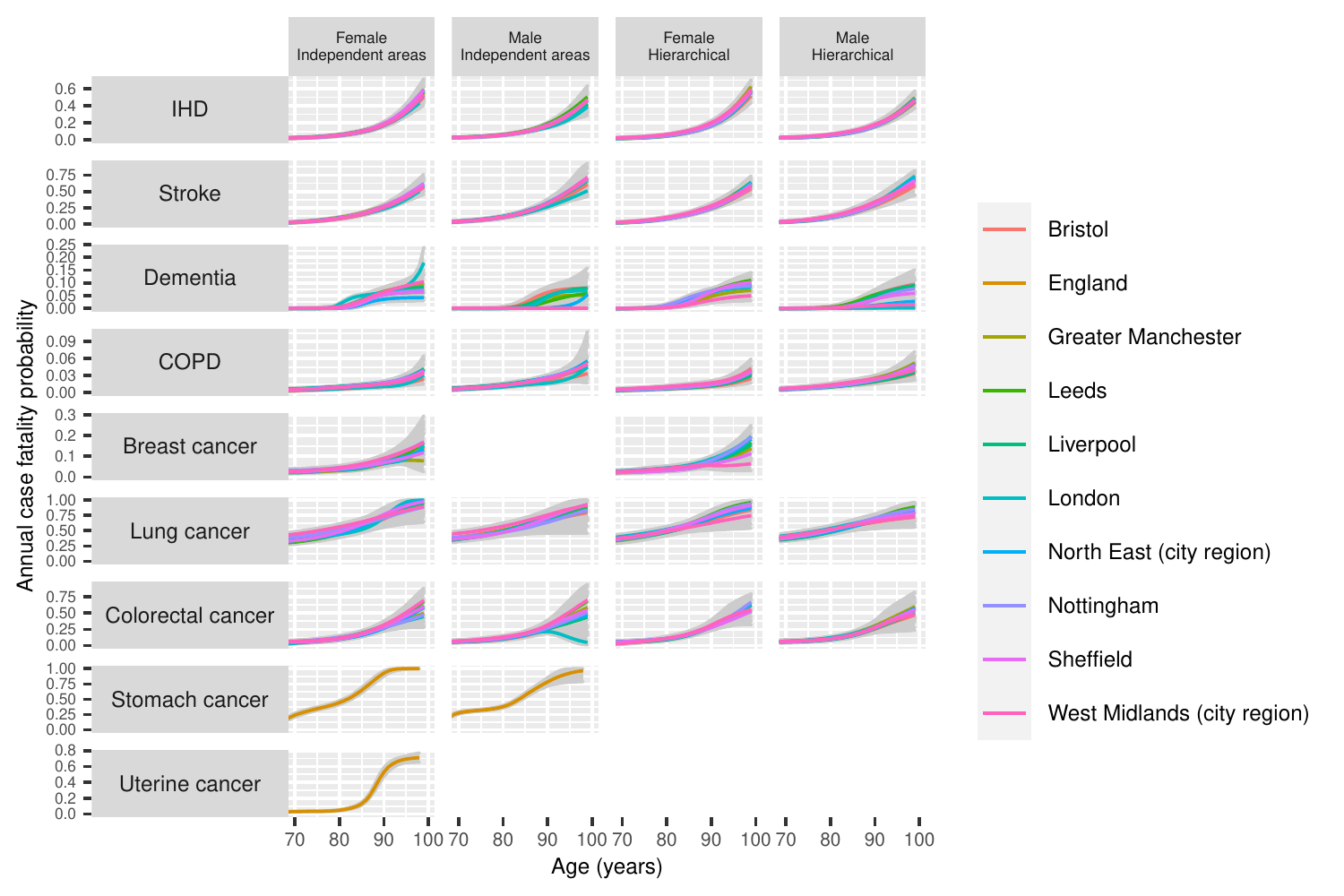}
  \caption{~Estimates of the case fatality probability for chronic diseases, compared between men and women, and comparing hierarchical models with models that treat city regions independently.  Each panel includes 9 lines denoting the posterior median from each of the 9 city regions, and the shaded area encompasses the 95\% credible intervals from all city regions.  For stomach and uterine cancer, only national estimates are produced from this model.}
  \label{fig:cf}
\end{figure}

For each disease, the posterior median and 95\% credible intervals for the annual case fatality probability, $1 - \exp(-i_a)$, obtained by MCMC under this model, are illustrated in the first two columns of Figure~\ref{fig:cf}.  Uncertainty is high beyond age 90 for most diseases, and the extent of between-area variability at younger ages is highest for lung cancer.

The approximation to the posterior around its mode is compared to the posterior from full MCMC in Appendix Figure 1, for case fatality estimates for multiple diseases, from women of a range of ages, in a single area.   The intervals produced by normal approximation around the mode generally agree with the intervals from the quantiles of the MCMC sample, while the posterior mode and median are similar.  This shows that a principled point estimate, together with a reasonable characterisation of the extent of uncertainty, can be produced without the expense of MCMC sampling (which took about 5 minutes per disease and patient group in this example, compared to 1 second for the approximation).

\subsection{Hierarchical models}
\label{sec:app:hier}

The variability between these area-specific estimates may be driven by noise, particularly where the uncertainty in the data is greatest for ages over 90 (Figures~\ref{fig:data},\ref{fig:cf}).   In these cases, a hierarchical model can allow information to be shared between areas, without assuming the risks are equal between areas.  The hierarchical model from Section~\ref{sec:hier} was implemented while also supplementing the data from 9 city regions of England with data from the rest of England.  Since sufficient direct data on incidence are available for each area, the random effects model was only used for case fatality, rather than incidence.   A mutually-exclusive set of 17 areas covering the whole of England was defined by (a) 9 city regions, (b) the regions not containing these city regions, and (c) the regions containing these city regions but with the city-region data excluded, using the lookup table at \url{https://geoportal.statistics.gov.uk/datasets/0c3a9643cc7c4015bb80751aad1d2594/explore}.

The hierarchical models were more difficult to develop, due to the complexity of the posterior distributions and the computational expense.  For all diseases, stable estimation required the parameters $\lambda_0$ and $\lambda_0^{(inc)}$, describing the smoothness of the rates as functions of age, to be fixed.  These were set to plausible values determined from the non-hierarchical models.   For uterine and stomach cancer, arbitrarily strong restrictions were required to enable between-area variations in case fatality to be estimated, specifically, an assumption that case fatality was a constant or log-linear function of age.  For these less common diseases, we would judge that precisely capturing area-level differences is not essential for informing health impact models for physical activity interventions, therefore we present only country-level estimates.

In general, the hierarchical and non-hierarchical models gave substantively the same estimates, except for the oldest ages where the data are sparsest.  At those ages, the variability between the area-specific estimates, and the extent of uncertainty, is less than under the model where areas are treated independently, since each area's estimate is ``shrunk'' towards the data from other areas, illustrated in the third and fourth columns of Figure~\ref{fig:cf} .   The cross-validatory statistics $\sum_i elpd_i$, however, indicated that the non-hierarchical models had better predictive ability on the whole (Appendix Table 2).

\begin{figure}[htbp]
  \includegraphics[width=\textwidth]{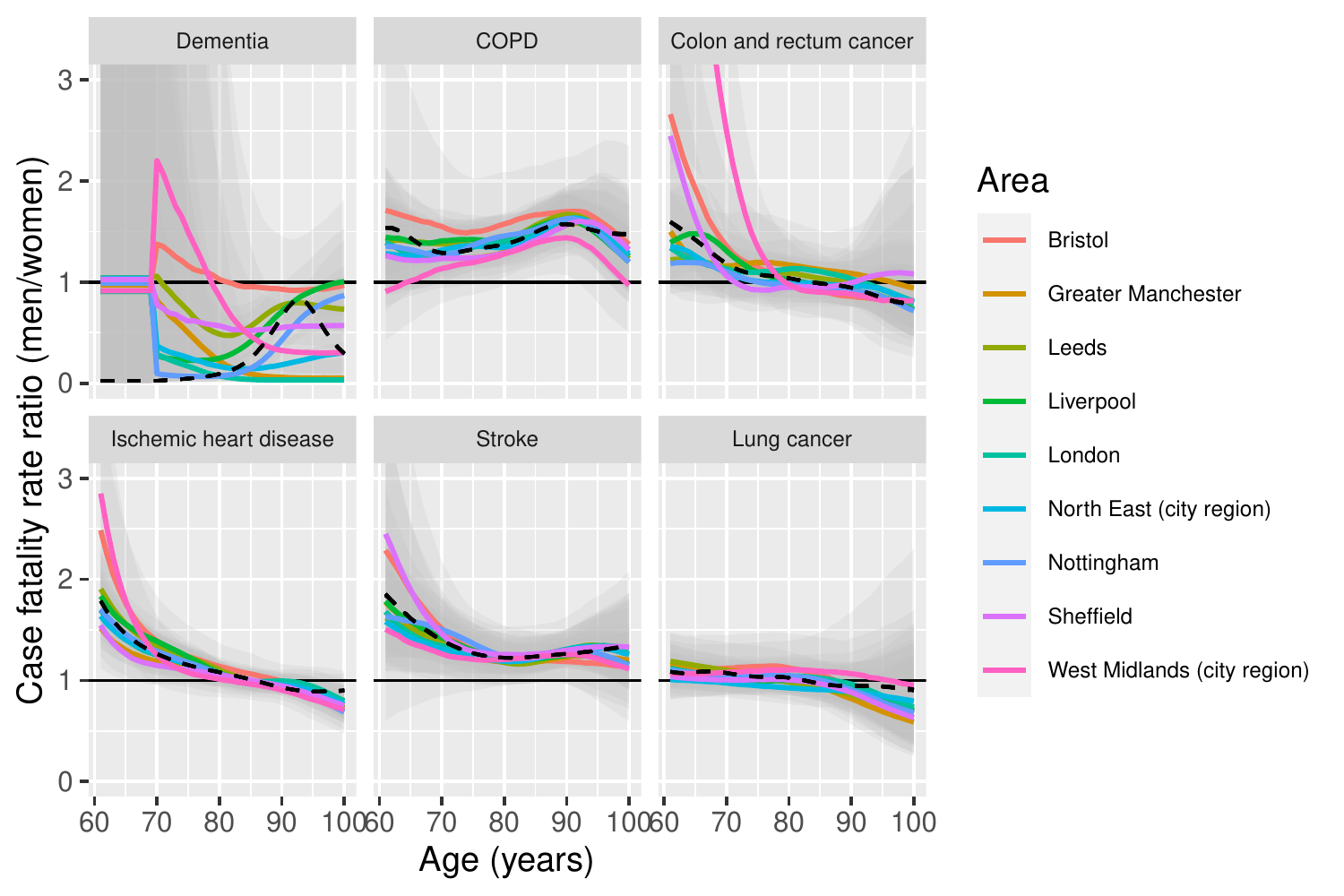}
  \caption{~Ratio of case fatality rate between men and women, for seven diseases, by age and city regions of England.  Posterior medians from the hierarchical model as coloured lines, with 95\% credible intervals in faint grey.  Posterior median from the hierarchical model with gender effect independent of area shown as a black dotted line.  Horizontal axis restricted to exclude the estimates with large uncertainty.}
  \label{fig:gender}
\end{figure}

Figure~\ref{fig:gender} illustrates the effect of gender.  The ratio of case fatality rates between men and women, estimated from the hierarchical models, is shown as a function of age, by disease and area.   For most diseases, the shape of the age-dependence of this gender effect generally appears similar between areas, with any differences in the estimated shape explainable by uncertainty from sparse data (e.g. colorectal cancer for around age 70 and younger).  For dementia, there is a large variability between these estimates that is driven by the small numbers of deaths by gender and area, in particular for men.   Therefore another set of hierarchical models was fitted, with the additional assumption that the relative case fatality between women and men was the same in every area (Section~\ref{sec:gender}), i.e. area and gender effects are additive.   The posterior median of the equivalent area-independent gender effect from this model is superimposed on Figure~\ref{fig:gender}.    For all diseases, apart from colorectal cancer and ischemic heart disease, the cross-validatory information criterion $-2 \sum_i elpd_i$ was lower for the model with additive gender and area effects (Appendix Table 2), indicating that it provided a better overall description of the data for that disease than the model where the gender effect interacts with the area effect.  For dementia, the additive model appears to be under-smoothing the effect of gender around ages 70-75.

\subsection{Time trends}

The model with areas treated independently was extended to include time trends for ischemic heart disease. Including time trends in the hierarchical model was computationally infeasible.   Estimates of trends over time in the incidence of myocardial infarction and subsequent case fatality (in terms of 30-day survival) are obtained from three sources: \citet{bhftrends2011} (for 1968--1998 in Oxfordshire, England), \citet{smolina2012determinants} (for 2002--2010 in England) and \citet{bhf2020} (for 2010--2019 incidence in the UK).  These data are illustrated in Appendix Figure 2.  No relevant data on case fatality after 2010 were found --- instead the trend from 2003--2010 was assumed to continue from 2010 to 2019.  Note that each source presents estimates by a different age grouping, leading to some inconsistency in the age-specific trends.    Estimates from 5-year age groups are converted to single years of age by smoothing, the 1968 values are assumed to apply to previous years, and the 1998-2002 values are interpolated, to obtain a matrix describing the ratio of incidence (or case fatality) rates between each calendar year and the current year (2019).  These ratios are used in the model from Section~\ref{sec:trends}, assuming that they apply to the incidence and case fatality for all ischemic heart disease, and that the trends apply identically to all areas of England.   To assess sensitivity to the assumption that the trend of reduction in case fatality beyond 2010 was the same as in the previous decade, we investigated an alternative scenario where the 2010 rate remained constant in subsequent years.

\begin{figure}
  \includegraphics[width=\textwidth]{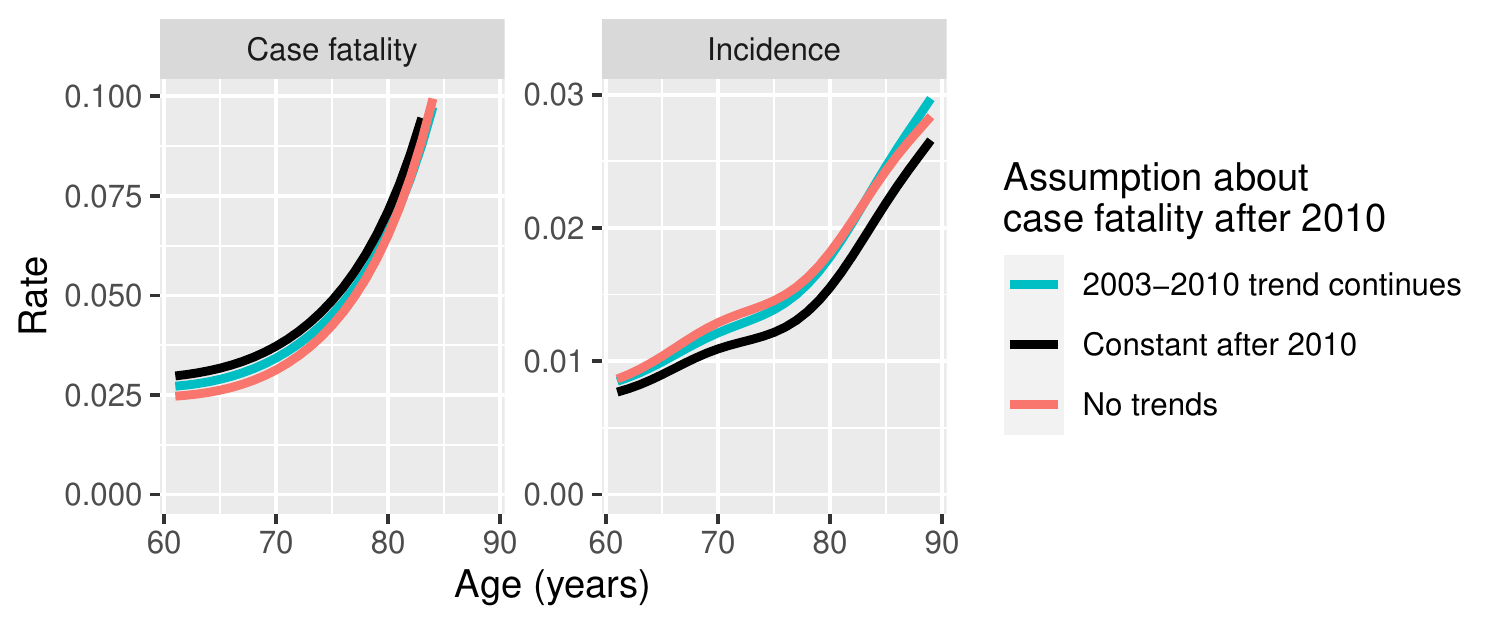}
  \caption{~Estimates of case fatality and incidence for ischemic heart disease by age, for men in the Leeds city region in 2019.  A model with no time trends in rates is compared with two models with time trends and different assumptions about case fatality after 2010.}
  \label{fig:ihd_trend_res}
\end{figure}

Estimates of case fatality and incidence for ischemic heart disease in 2019, for an example area (Leeds city region), by age and gender, are plotted in Figure~\ref{fig:ihd_trend_res}, comparing the model that ignored time trends with the models that included the estimates of past trends in these risks.  Two assumptions about the trends in case fatality after 2010 are compared.  Estimates of case fatality under age 80 are slightly higher, and estimates of incidence slightly lower, when these trend estimates are included.  The differences are more substantial under the assumption where case fatality does not change after 2010.

\citet{scarborough2016assessing} estimated a similar size and direction of bias for estimates of IHD incidence from prevalence data, when the Dismod II model is used and the time trend is ignored.  Note that our model extends the model in DisMod II to allow the time trend to vary with age.   In general, it is difficult to predict how accounting for a specific past trend will change estimates of current incidence or case fatality rates from current mortality and prevalence data.   The current disease outcomes depend on both the current incidence and case fatality and the current prevalence (equation~\ref{eq:mort}) which depends on the risks in previous years for the people represented in the current data.  If these people faced higher risks of both incidence and case fatality in the past, the effect on the current prevalence (and hence our estimates) is hard to predict.   The model can enable sensitivity analysis in situations where the past trends and their influence are uncertain.

\begin{figure}
  \includegraphics[width=\textwidth]{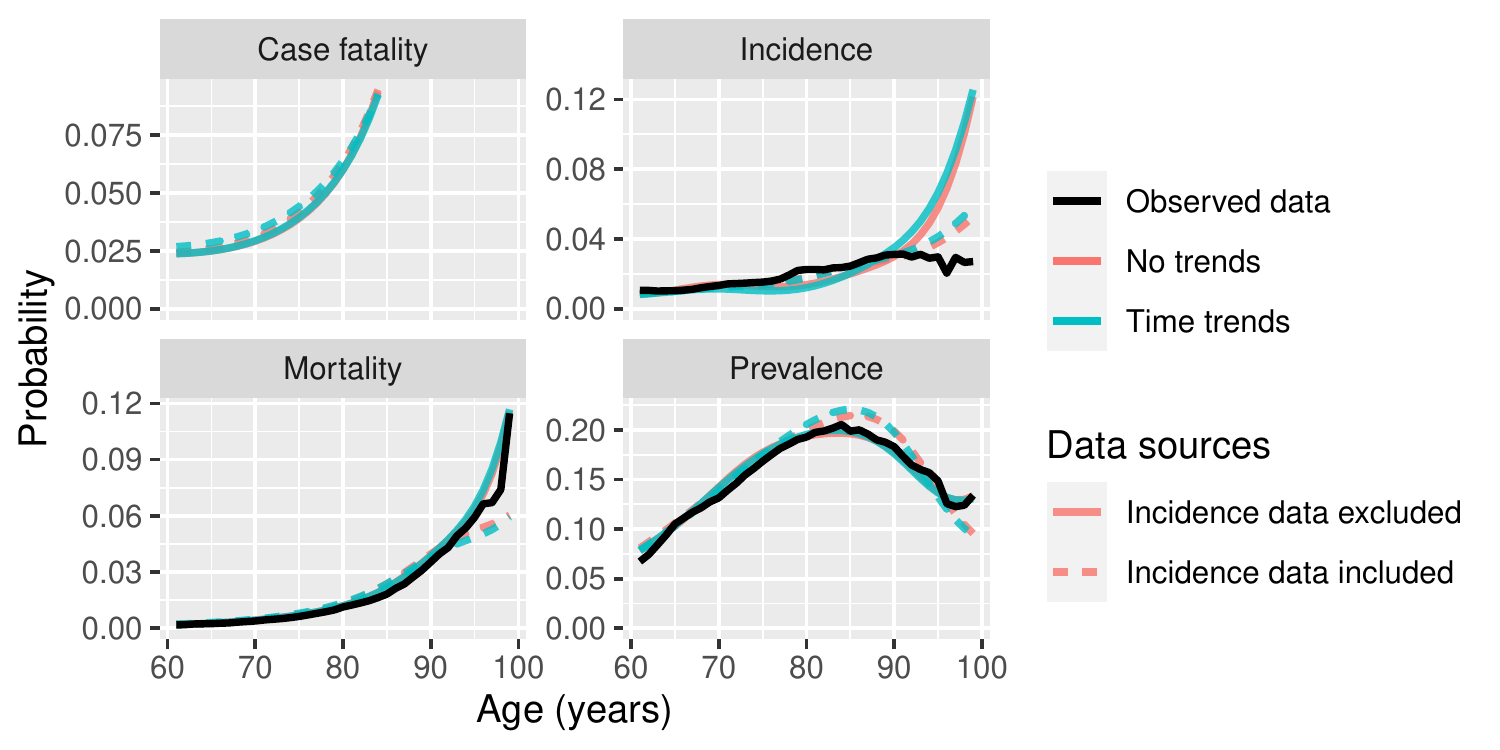}
  \caption{~Estimated (posterior mode) case fatality, incidence, prevalence and mortality probabilities for ischemic heart disease for men in the Leeds city region, compared with corresponding observed proportions for all of these measures except case fatality, which is unobserved.  A model with and without time trends is compared, and a model (without time trends) fitted to incidence, prevalence and mortality data is compared with a model fitted to just prevalence and mortality data (excluding incidence).}
  \label{fig:checkfit}
\end{figure}

\subsection{Data consistency and goodness of fit}
\label{sec:res:fit}

Finally we show how the models can be checked against observed data, and show the influence of different data sources on the estimated rates.  Again we select a single disease, area and gender for illustration.   The model for ischemic heart disease with no time trends is compared against the model with time trends (assuming trends continue after 2010), for men in the Leeds city region.   Additionally we fit two further models, with and without time trends, in which the incidence data were excluded, hence case fatality and incidence rates are inferred from current prevalence and mortality data alone.

The observed data (in the form of proportions) are compared against posterior mode estimates of the theoretical probabilites assumed (Section~\ref{sec:homog}) to generate those data.   Specifically we compare
\begin{itemize}
\item $y_a^{(inc)}/n_a^{(inc)}$ with estimates of the theoretical annual incidence probability $1-P_{a11}$

\item $y_a^{(prev)}/n_a^{(prev)}$ with estimates of the theoretical prevalence $\pi_a$

\item $y_a^{(mort)}/n_a^{(mort)}$ with estimates of the theoretical annual mortality probability $d_a$
\end{itemize}
In the models with time trends, the theoretical probabilities are based on the 2019 rates, to match the year of the data, but here the year subscript in the notation of Section~\ref{sec:trends} is suppressed for clarity. 

The estimates from the models that combine all three sources of evidence are shown as dotted lines in Figure \ref{fig:checkfit}.  When the incidence data are included in the fit, there is some disagreement between the observed and fitted incidence data, and between the observed and fitted mortality and prevalence data beyond around age 80.  When the incidence data are excluded, the fit to the incidence data is worse.   This suggests that the incidence and prevalence data provide conflicting information about incidence rates.  The estimates disagree even if a time trend is included, suggesting the inclusion of time trend data is insufficient to explain this conflict.  This may either because the time trend data used here do not accurately represent the trends in risk experienced by the population behind the GBD data, or that the incidence data represent a different population from the prevalence and/or mortality data.  The inferred case fatality rates only change moderately when the incidence data are excluded, however. 

While the point estimates disagree slightly, the associated uncertainty (not plotted) is large enough to suggest that any conflicts are not statistically significant.  This was tested formally by computing the conjugate posterior distribution for the probability $p$ underlying each observed proportion, by combining, e.g. $y_a^{(inc)} \sim Bin(n_a^{(inc)}, p)$ with a Beta(0.5, 0.5) prior.  The posterior probability $q$ that $p<p^{(full)}$ (where $p^{(full)}$ is the comparable quantity from a synthesis of indirect evidence) is then computed to obtain a two-sided \emph{conflict p-value} $2 \min(q,1-q)$ \citep{presanis2013conflict}, which is $<0.05$ in less than 5\% of all cases, favouring the hypothesis that $p=p^{(full)}$.

If conflict is a concern, then the incidence data could be excluded to ensure that the inferred prevalence and mortality match the corresponding observed data.  Therefore, to maintain a transparent connection between the inferences and the data, while minimising assumptions about time trends and similarity between populations, the case fatality estimates could be taken from the prevalence and mortality data alone, while taking incidence estimates from the incidence data alone.  Intuitively, current prevalence and mortality data are more informative about the current case fatality rate (which describes the risk of death for the population with prevalent disease) than current incidence rates.  In Equation \ref{eq:mort}, $P_{a13}$, the risk of both getting the disease and dying from it within a year, will generally be much smaller than the case fatality probability $P_{a23}$, suggesting that the data on prevalence $\pi_a$ and mortality $d_a$ are more important for learning $P_{a23}$ than data on incidence.

Finally, the non-hierarchical models from Sections~\ref{sec:app:homog} were fitted again for all diseases with the incidence data excluded.  (Without the direct data on incidence, the MCMC samplers for the hierarchical models failed to adapt and converge to the posterior distribution within a day of run time, even with hyperparameters fixed.). These were seen to fit the data slightly better on average than the models excluding incidence data, though both fitted adequately and the difference in their estimates was small (see Appendix Section E.1).

\section{Discussion}
\label{sec:discussion}

This paper has provided more principled, transparent and flexible methodology, and accessible software, for estimating disease incidence and case fatality rates given indirect data.  The methods were motivated by the requirements of multistate lifetable models to assess the health impacts of interventions or scenarios for prevention of chronic diseases.   The methods may also be more widely useful for describing the burden of disease in settings where only indirect data are available, as has previously been done in the Global Burden of Disease project.  We have illustrated a range of models of different complexity and computational expense.    In principle, the Stan software that was used would enable variations and extensions of these models to be programmed in the same way, though in practice, more complex models may be more difficult to identify from the data.  More complex models may also be more computationally intensive, and MCMC estimation for complex models can be particularly challenging where the data are weaker, as we found for the less common cancers.  However, we have shown how optimisation can be used to provide point estimates of rates, and approximate credible intervals, without the expense of MCMC.

The required model complexity depends on the purpose of the model.  For describing the burden of disease in different populations, it may be helpful to extend the model to allow more detail, e.g. to capture variations between subgroups other than by age and gender (using parametric assumptions as in Section~\ref{sec:gender}), to describe spatial correlations between smaller areas, or to use covariate information to strengthen prediction at smaller areas.  In contrast, if the ultimate aim is to estimate population health impacts of, e.g. changes in physical activity or diet, then simpler models will often be sufficient, since detail is only required if it would affect the estimates of impact.  For example, for a rare disease, the information about case fatality will automatically be weaker, however impacts on a rare disease will only form a small part of the population impacts.  Thus it may not be worth capturing variations more precisely by a more structured model, unless the specific impacts on that disease are of interest.

Ideally there would be no need for models, and instead, disease burden would be estimated from direct data.   While uncertainty in our model-based estimates can be quantified, models rely on structural assumptions, hence these uncertainties are likely to be understated.   These assumptions include the approximation of disease as a three-state Markov process, which may conceal variations between individuals and through time.  For many diseases, there are risk prediction models (e.g. QRisk,~\citet{qrisk3}) that give better descriptions of how incidence varies with individual characteristics, and these can be used as part of microsimulation models for estimating health impacts~\citep{mytton2017modelled}.   Case fatality, however, is less well understood. The Markov assumption, that case fatality does not depend on time since onset, is unrealistic for some diseases.  Commonly it is higher soon after onset, when acute events, such as myocardial infarction, may have occurred.  Case fatality rates are also affected by treatments and the presence of other diseases (multimorbidity).  While violation of these assumptions is a concern if case fatality rates are of direct interest, it is less clear how much these assumptions would affect estimates of the impacts of prevention interventions.  Interaction between diseases might be represented in a multi-state model by introducing ``combined'' disease states \citep{lauer2003popmod}, though the computational difficulty and data requirements would increase rapidly with the number of diseases.  Another approach to health impact modelling is to disregard causes of death and model the all-cause mortality rate for people with each disease separately, see, e.g. \citet{boshuizen2017taking}. 

In our application, we determined case fatality and incidence estimates, for several diseases assumed to be relevant to active transport, relating to city regions in England, to inform multistate health impact models.  For ischemic heart disease, one of the diseases most affected by physical activity, we extended previous methods to show how including information about past trends in age-specific risks can have a substantial impact on estimates of current risks.    We illustrated the uncertainty in estimates that arises from conflicting information in different sources of data, and uncertainty about past time trends.    Note also that the Global Burden of Disease estimates that we used as data for our case study are likely themselves to have been derived from models, but the full details of how they are derived are unclear.    These uncertainties show that better data is needed on case fatality rates for chronic diseases, with transparent information about how the data were generated.  This would provide better information about disease burden, and enable better-informed estimates of the impacts of interventions to prevent disease.

\paragraph{Acknowledgements}
CJ was supported by the Medical Research Council, programme number MRC\_MC\_UU\_00002/11.  This project (JW and CJ) received funding from the European Research Council (ERC) under the Horizon 2020 research and innovation programme (grant agreement No 817754). This material reflects only the author's views and the Commission is not liable for any use that may be made of the information contained therein.  JW and CJ were also supported by the METAHIT project (MRC Methodology Panel MR/P02663X/1).  BZD is supported by a RMIT VC fellowship.  The authors are grateful to Daniela De Angelis for helpful comments.

\appendix
\maketitle
\section{Transition probabilities in terms of rates}
\label{sec:kolmogorov}

The disease model described in section~\ref{sec:model} of the main manuscript is a continuous-time Markov model with the transition intensity matrix at age $a$ years:
\[
Q_a = \left(\begin{array}{lll}
-i_a  & i_a  & 0 \\ 
r_a   & - (i_a + r_a)  & f_a  \\ 
0 & 0 & 0
\end{array}\right)
\]
The corresponding \emph{transition probability matrix} $P_a(t)$ is the matrix of probabilities $p_{a,j,k}(t)$ that a person in state $j$ at an age of exactly $a$ years is in state $k$ at a point $t$ years later.   This is the solution to the Kolmogorov forward equation $\frac{d}{dt} P_a(t)= P_a(t) Q_a(t) $, with $P_a(0)$ equal to the identity matrix \citep[see, e.g.][]{cox:miller}.   For a person aged exactly $a$ years, the annual transition probabilities $P_a = P_a(1)$ governing their state at age $a+1$ can be obtained as a closed-form function of the incidence $i_a$, case fatality $f_a$ and remission $r_a$ rates at that age, since these rates are assumed to be constant within a single year of age.  As described in \citet{dismod2}, and dropping the $a$ subscript for clarity, the $j,k$ entries $p_{j,k}$  of this matrix are 
\begin{eqnarray*}
  \label{eq:1}
p_{1,1}  & = &  (2(v-w)(f+r) + v(q-u) + w(q+u)) / (2q) \\
p_{1,2}  & = &  i(v - w)/q \\
p_{1,3}  & = &  (-u(v-w) - q(v+w))/(2q) + 1 \\
p_{2,1}  & = &  (v-w)r/q \\
p_{2,2}  & = &  -((2(f+r) - u)(v-w) - q(v+w)) / (2q) \\
p_{2,3}  & = &  ((v-w)(2f - u) - q(v+w))/(2q) + 1\\
p_{1,3}  & = & 0 \\
p_{2,3}  & = & 0 \\
p_{3,3}  & = & 1 \\
\end{eqnarray*}
where state 1 is no disease, state 2 is disease,  state 3 is death, $u = i + r + f$,  $q = \sqrt{i^2 + 2ir -  2if  + r^2 + 2fr + f^2}$, $w = \exp(-(u + q) / 2)$ and $v = \exp(-(u - q) / 2)$. 

\section{Obtaining count data from published information} 
\label{sec:propstocounts} 

Disease epidemiology summaries might be published as proportions rather than counts.    For example, prevalence may be published as an estimated proportion of people with the disease $p = \hat{p}_a^{(prev)}$.    To convert this into a count $r$ and denominator $n$, as required for the model in Section~\ref{sec:homog} of the main manuscript, additional information is required (dropping the age subscript in this section for clarity).    Most simply, if the population size $n$ that informed the estimate is published, or can be confidently estimated, then we determine the count as $r = \hat{p} n$. 

If the population size is unknown, but a interval estimate $(\hat{p}^{(L)},\hat{p}^{(U)})$ (95\%, say) is published that represents uncertainty about the point estimate $\hat{p}$, then this information can be converted to an implicit numerator $r$ and denominator $n$ with a Bayesian technique used in expert elicitation \citep{ohagan:elic}.    This assumes that the point and interval estimates are summaries of a Beta posterior distribution which has been obtained by combining a vague prior with an observation of $r$ events occurring out of a sample of $n$ individuals.   For example, with a vague $Beta(0,0)$ prior (which is uniform on the logit scale) the posterior is $Beta(r, n-r)$.  We can then search for the best-fitting $Beta(\alpha, \beta)$ distribution which has median $\hat{p}$ and (2.5,97.5) quantiles $(\hat{p}^{(L)},\hat{p}^{(U)})$, and set $r=\alpha, n=\alpha+\beta$.  
This is done by minimising the following expression with respect to $\alpha$ and $\beta$ using numerical optimisation.  This expression is the sum of squared errors between the published  
quantiles and the theoretical quantiles based on the Beta distribution,
\[  
\left(F(\hat{p}^{(L)} | \alpha,\beta)  - 0.025\right)^2 + 
\left(F(\hat{p}  | \alpha,\beta)  - 0.5\right)^2 +
\left(F( \hat{p}^{(U)} | \alpha,\beta)  -0.975\right)^2 
\] 
where $F$ is the cumulative distribution function of the Beta$(\alpha,\beta)$.

If we also suspected that one of the data sources may be biased, but were unsure about the direction of bias, we could downweight that data source by multiplying both $r$ and $n$ by the same amount and rounding, e.g. by 0.5 if we wanted to give a data source half its original weight. 

\section{Handling data coarsened by age groups} 
\label{sec:agedisagg}

A further complication of typical chronic disease burden data is that it may be published as estimates by broad age groups, e.g. 5 or 10-year ranges, possibly of unequal widths, while the multistate disease model of Section \ref{sec:model} of the main manuscript requires data for each year of age.  \citet{flaxman2015} discuss a variety of methods for modelling rates published on unequally-spaced age groups, but acknowledge computational or statistical limitations with them all.  In DisMod II, for example, rates for coarse age groups can be converted to smooth functions of age by polynomial or spline interpolation, however that is not guaranteed to preserve the underlying information that the estimates were obtained from. 

Instead, we use \emph{smooth temporal disaggregation}, applied to the count data, rather than the underlying rates.  For example, given a published prevalence count of $\sum_{a=51}^{55}{y_a}^{(prev)}$ aggregated over ages 51--55, counts per year of age could be produced naively by dividing the 5-year total by 5.  However, \emph{smooth} disaggregation methods produce more realistic estimates of ${y_a}^{(prev)}$ that form a smooth function of $a$ over all age intervals, while preserving the sum within each age interval to exactly equal the published count.   A variety of methods for doing this efficiently are implemented in the \texttt{tempdisagg} R package \citep{RJ-2013-028}.   The count $y$ and denominator $n$ for each different disease metric are disaggregated separately.

\section{The `disbayes' R package} 
\label{sec:disbayes}

An R package, \verb+disbayes+, was developed, that implements all the models used in this paper.   It is available from \url{https://chjackson.github.io/disbayes}.  A brief example of its use and a description of its features is given here, but full documentation is available as R help pages for the functions in the package, and in a ``vignette'' containing worked examples.

\subsection{Non-hierarchical models} 
The user must supply a data frame with one row for each year of age (beginning at year 0 but ending at any year), for a homogenous population (e.g. as defined by a single area and gender).   The columns contain estimates of at least mortality and at least one of incidence or prevalence, and optionally also remission.    The recommended form is as numerators and denominators, as described in Section~\ref{sec:homog} of the main manuscript, so that rows of the data might look like this: 

{\small
\begin{verbatim}
    age gender area  inc_num inc_denom prev_num prev_denom mort_num mort_denom ...
1    60 Male   Leeds      55      5252      180       2636       20      12386
2    61 Male   Leeds      64      6157      168       2217       21      11978
3    62 Male   Leeds      68      6437      163       1901       22      11700
4    63 Male   Leeds      65      6095      163       1686       24      11553 ...
\end{verbatim}
}

Alternatively, point and interval estimates, or point estimates and denominators, can be provided.  The heuristic explained in Appendix Section~\ref{sec:propstocounts} is used internally to convert credible intervals to numerators and denominators.  The important requirement is that as well as an estimate, some indication of sampling uncertainty must be provided, either through a denominator or a credible interval.  This enables formal statistical inference.

If data are available as coarsened age intervals, this must first be converted to counts per year of age, e.g. using the methods discussed in Appendix Section~\ref{sec:agedisagg}.  A worked example of this is provided in the package vignette. 

The model in Sections \ref{sec:homog}--\ref{sec:age} of the main manuscript is then fitted with a command such as the following 
\begin{verbatim}
db <- disbayes(data = dat, 
               inc_num = "inc_num", inc_denom = "inc_denom",
               mort_num = "mort_num", mort_denom = "mort_denom",
               prev_num = "prev_num", prev_denom = "prev_denom",
               eqage = 30)
summ <- tidy(db) 
\end{verbatim}
where the argument \verb+dat+ gives the name of the data frame, the following six arguments name the variables in that data giving numerators and denominators for incidence, mortality and prevalence, and \verb+eqage=30+ specifies that case fatalities are assumed to be equal for all ages below 30 (Section~\ref{sec:age} of the main manuscript).  

Additional options to the \verb+disbayes+ function include:
\begin{itemize}
\item \verb+cf_model+: model for case fatality as a function of age. By default, a smooth function of age is used, \verb+(cf_model="smooth")+, with a shape that is unrestricted except through \verb+eqage+.  Alternative options include a smooth increasing function \verb+("increasing")+, a model where rates are constant with age (\verb+"const"+), or unrestricted rates that are estimated independently for each age (\verb+"indep"+).   Similar options are available to specify the model for incidence or remission rates. 
  
\item \verb+method+: The algorithm used to fit the model.  The default \verb+"opt"+ uses optimisation to find the posterior mode, and obtains posterior credible intervals through a normal approximation.  The alternative \verb+"mcmc"+ uses MCMC sampling from the full posterior, which is more accurate, but an order of magnitude slower.  \verb+rstan+'s variational Bayes methods can also be used, and may give a more accurate approximation to the posterior than \verb+method="opt"+ for not much more expense, but they have not been investigated in detail for this class of models.

\item Options to control sampling or optimisation can be passed through \verb+disbayes+ to the underlying functions used by \verb+rstan+.

\item Options are available to specify the parameters of all prior distributions.

\item \verb+cf_trend+, \verb+inc_trend+: matrices of constants describing trends through time in rates, in the form of the ratio of case fatality (or incidence) between previous years and the year of the data, by year of age. 

\item \verb+disbayes+ can also return leave-one-out cross-validation statistics that describe how well each observation in the dataset would be predicted if the model were fitted to the remainder of the data, defined and computed using the methods of~\citet{loocv,loor}. 

\item \verb+hp_fixed+.  Sometimes there may be difficulties in obtaining the posterior distribution of the ``hyperparameters'' $\lambda_0,\lambda_0^{(inc)}$ that determine the smoothness of the function relating case fatality or incidence rates to age.  In those cases, the $\lambda_0$ can be fixed at values supplied in this argument, obtained, e.g. from an estimate in a similar dataset, or from the posterior mode if that can be determined.  Values of around 1--5 were estimated in the applications in the paper.  Higher values give more flexible curves, and values close to zero approach a linear function for the log rate with age.
\end{itemize}

The object \verb+summ+ then contains point and interval estimates for all estimated quantities in the model as a ``tidy'' data frame, a form convenient for processing and plotting, e.g. by using the \verb+dplyr+ and \verb+ggplot2+ packages.  For example, to extract from this data frame the posterior mode and approximate 95\% credible limits for the case fatality rates for people aged from 61 to 65, 
\begin{verbatim}
> library(dplyr)
> summ %>% 
>    filter(var=="cf", between(age, 61, 65)) %>% 
>    select(age, mode, "2.5%", "97.5%")
  age       mode       2.5%      97.5%
2  61 0.01480355 0.01234622 0.01738378
3  62 0.01507347 0.01263041 0.01766254
4  63 0.01535594 0.01296970 0.01803932
5  64 0.01565776 0.01337025 0.01833570
6  65 0.01598639 0.01373186 0.01868637
\end{verbatim}

The variables that can be extracted from this data frame are documented in \verb+help(tidy.disbayes)+.

\subsection{Hierarchical models} 

The function \verb+disbayes_hier+ can be used in a similar fashion to fit the hierarchical models of Sections~\ref{sec:hier}--\ref{sec:gender} in the main manuscript.  Case fatality rates can have random intercepts and slopes, but random effects are not currently supported for incidence or remission.   The dataset is supplied in the same form as above, but should describe data from multiple areas.  For example, if there are 100 years of age, and 10 areas, then the data should contain 1000 rows giving disease outcomes for each area and age.   Arguments to the \verb+disbayes_hier+ function include:
\begin{itemize}

\item \verb+group+: name of the variable in the data indicating the area or other group that the estimate was obtained from.

\item \verb+gender+: name of the variable indicating gender (or other binary classification).  This should be omitted if modelling a single gender.  If it is supplied, then gender and area effects are assumed to be additive, as in Section~\ref{sec:gender}, and the data should contain, e.g. 2000 rows if there are 100 ages and 10 areas.

\item \verb+hp_fixed+. A list of hyperparameters that should be fixed at supplied constants, or at their posterior modes.   These might include the smoothness parameters $\lambda_0$ or the standard deviations of random effects.  This might be used if there is difficulty sampling from a full hierarchical model that represents the posterior distributions of these parameters.
\end{itemize}

\subsection{Summary of packages}

The following table briefly summarises the differences between three alternative packages that are available for estimating the rates of the three-state disease model given indirect data on at least mortality, prevalence and/or incidence.  The main manuscript of this paper explains these differences in more technical detail.   A wide variety of useful models can be specified in any of these packages.  

It is assumed that the methods described in detail in \citet{flaxman2015} are those that are implemented in DisMod-MR, though the documentation included with the software itself is sparse.  Note also that DisMod-MR has more features for modelling \emph{direct} data such as prevalence data, including regression models to explain heterogeneity in disease burden between areas in terms of covariates.   DisMod II also has an extensive graphical user interface for interactive data management and graphics, however we would argue that due to their ``scriptable'' nature, languages such as R or Python are preferred for reproducible research. 

\begin{table}[h]
\begin{tabular}{p{1.2in}p{1.2in}p{1.2in}p{1.2in}}
\hline
& DisMod II & DisMod-MR & \texttt{disbayes}\\
\hline
Statistical principles &  Maximum likelihood & Bayesian & Bayesian \\
Age dependence of rates & Informal smoothing & Formal, fixed-knot penalised linear spline & Formal, more flexible penalised spline\\
Between-area variations in rates & Areas independent & Empirical Bayes & Fully-hierarchical or empirical Bayes \\
Time trends in rates & Age-common & None & Age-dependent \\
Regression models & No & Yes & No \\ 
Computational method & Optimisation (fast) & MCMC (intensive) & Optimisation or MCMC \\
Interface & Windows & Python & R \\
\hline
\end{tabular}
\caption{Summary of characteristics of three alternative programs for multistate disease modelling from indirect data}
\end{table}

\clearpage

\section{Supplementary data and results}
\label{sec:supp}
~
\begin{figure}[h]
  \includegraphics[height=5in]{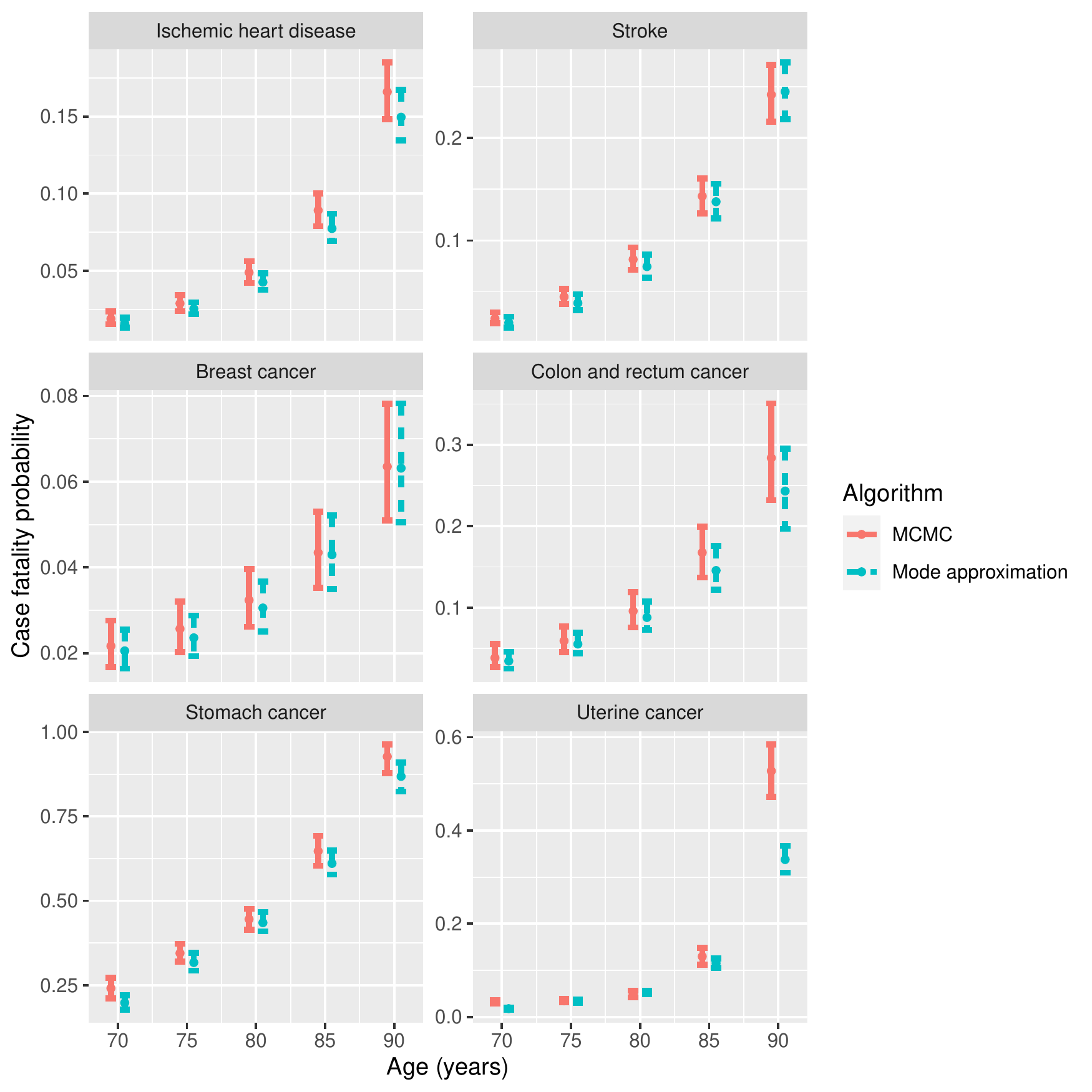}
  \caption{Case fatality probabilities for chronic diseases for women in Bristol (or, for uterine cancer and stomach cancer, in England), comparing posterior median and 95\% credible intervals obtained from a MCMC sample with posterior mode and 95\% credible intervals obtained from a normal approximation around the mode.}
  \label{fig:opt_compare}
\end{figure}
~
\begin{figure}[h]
  \includegraphics[height=2.6in]{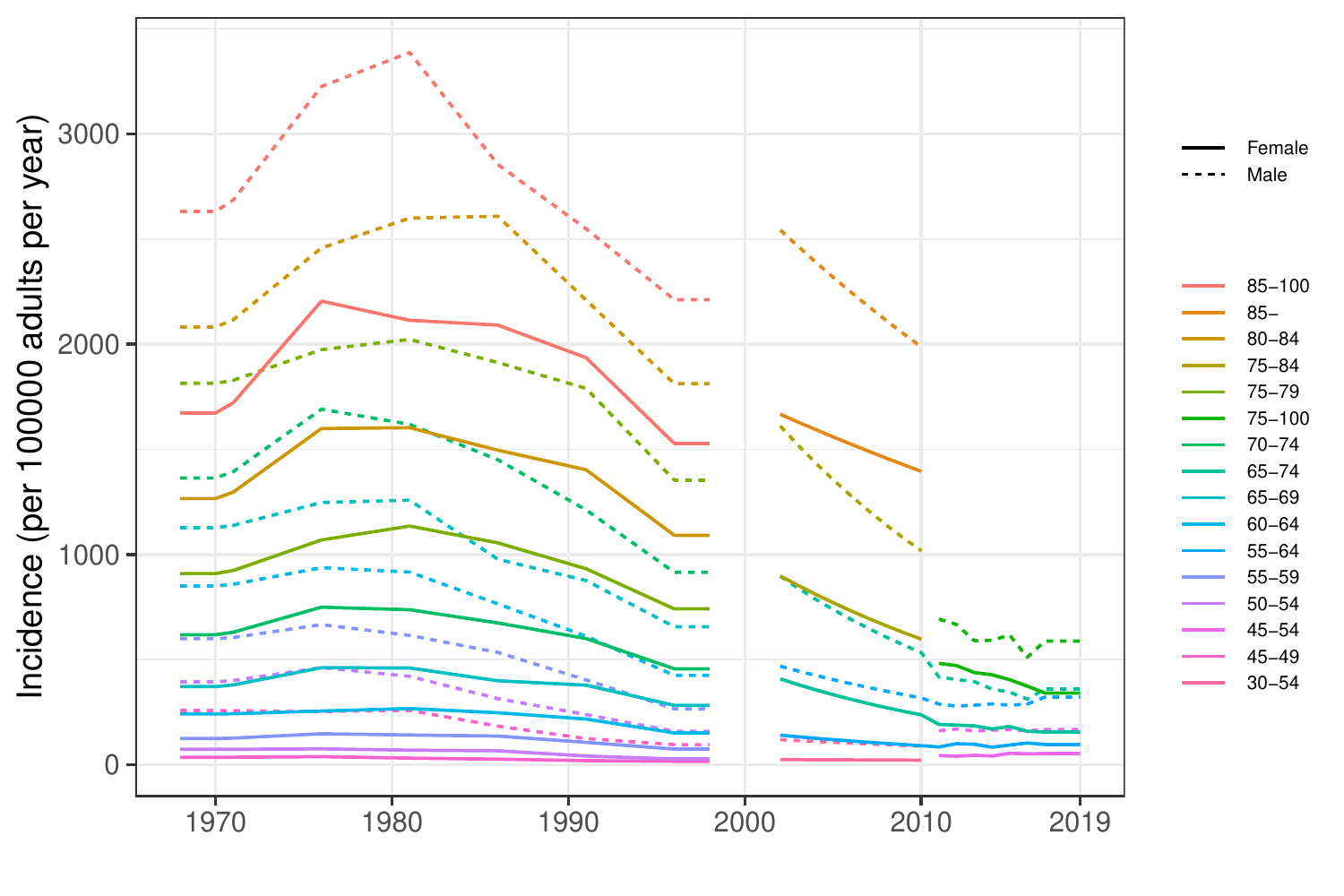}~\\
  \includegraphics[height=2.6in]{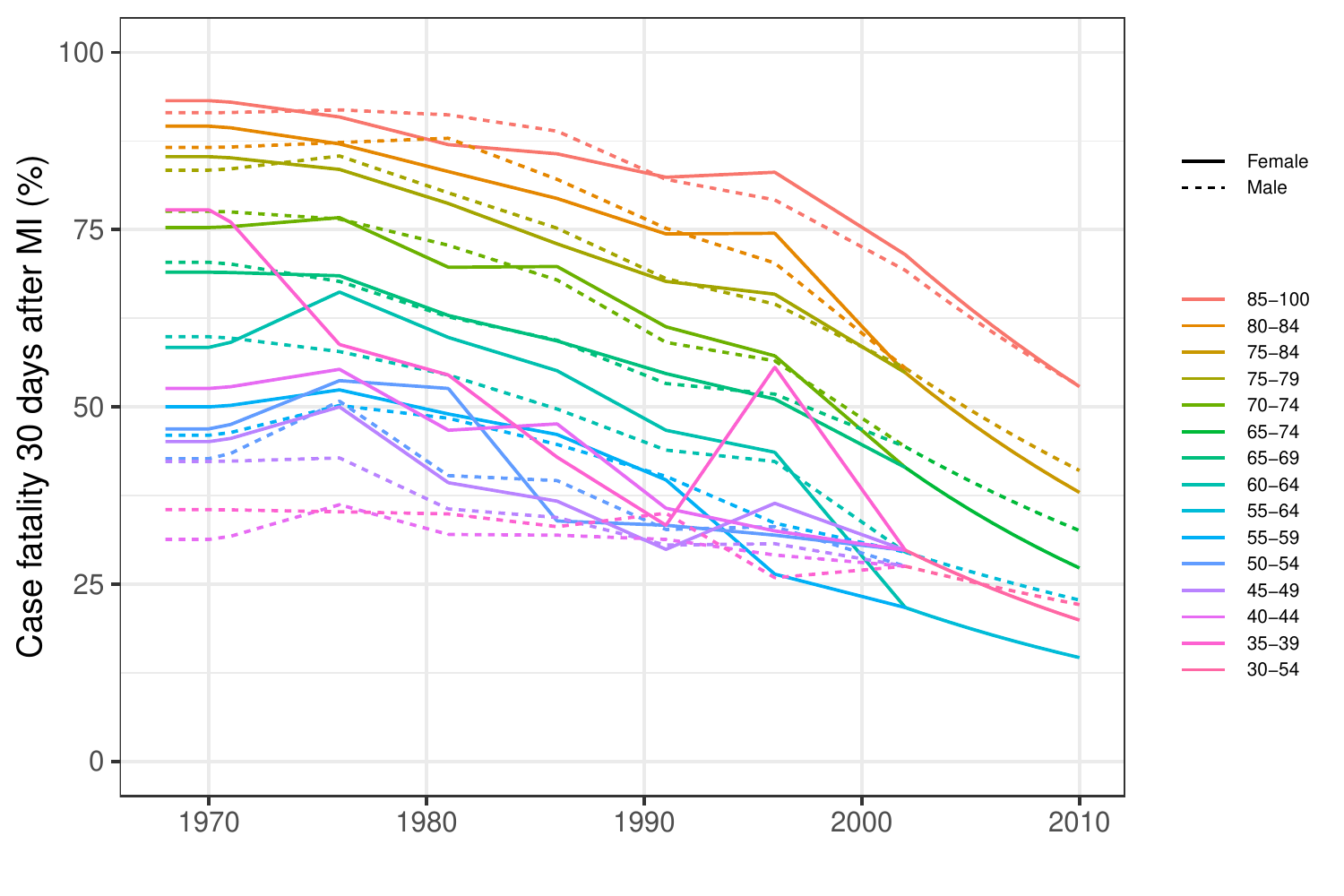}
  \caption{Estimated trends through time in the incidence (top) and case fatality (bottom) of myocardial infarction, by gender and age group \citep[from][]{bhftrends2011,smolina2012determinants,bhf2020} (values in identical age groups are interpolated).}
  \label{fig:ihd_trends}
\end{figure}

\subsection{Cross-validatory model comparison}
\label{sec:supp:cv}

For each of the outcomes for which we had area-specific data, that is, mortality, incidence and prevalence, the observation-specific expected log predictive density $elpd_i$ was computed via the Pareto-smoothed importance sampling method of~\citet{loocv}. This did not always give reliable estimates, since the importance sampling weights sometimes had an excessively high variance, as indicated by an estimated shape parameter $\hat{k}>0.7$ for the generalized Pareto approximation to the weights.  This happens when the full posterior has a much higher variance than the posterior after leaving out one observation, due to the left-out observation being particularly influential, in which case the full posterior is an unreliable basis for an importance sampling approximation. 

In our models, $\hat{k}<0.7$ consistently held for predictions of mortality between ages 50 and 90, but not for predictions of incidence or prevalence.  Since the main goal of the models is to predict case fatality for chronic diseases within these ages, we judge this to be an adequate basis for comparing the non-hierarchical model with the hierarchical models with and without the additive effect of gender.  Table~\ref{tab:cv} shows the results as differences in ``leave-one-out information criteria'' ($LOOIC=-2\sum_i elpd_i$) relative to the non-hierarchical model.  The non-hierarchical models are preferred in most cases, except for COPD, where the additive hierarchical model is preferred, since the relative risk between men and women is similar between areas (Figure~\ref{fig:gender} in the main manuscript).  The hierarchical model was also preferred for ischemic heart disease, though note that this model choice is of fairly minor importance, since the estimated case fatality probabilities differed by less than 0.01 between the models in over 90\% of ages and areas. 

Importance weights could not be computed accurately for the model without the incidence data (which is essentially based on a smaller dataset, hence more data-points are judged to be influential).  Checks of agreement of estimates with data are more helpful for comparing the models with and without the incidence data, as illustrated in Section~\ref{sec:res:fit} of the main manuscript.  Data as in Figure~\ref{fig:checkfit} of the main manuscript can be summarised numerically for all diseases and areas, e.g. the mean absolute difference between the observed mortality proportions and fitted mortality probabilities is 0.0007 for the model with incidence data, and by 0.0002 for the model excluding incidence data.  Therefore, while excluding incidence data gives a slightly better fit, the agreement with the data is good for both models.  Estimates of the case fatality probability under these two models differ by less than 0.01 for 90\% of the age, area, gender and disease-specific estimates -- therefore on the whole, these models do not severely disagree with each other.

\begin{table}[h]
  \begin{tabular}{p{1.5in}p{1in}p{0.7in}p{0.7in}}
\hline
  & Non-hierarchical model & \multicolumn{2}{l}{Hierarchical models}  \\
  &    &  Interaction  & Additive \\
    \hline
    Ischemic heart disease  &  0  &  -34  &  7\\
    Stroke  &  0  &  160  &  127\\
    Dementia  &  0  &  124  &  108\\
    COPD  &  0  &  23  &  -85\\
    Breast cancer  &  0  &  -2  &  \\
    Lung cancer  &  0  &  22  &  17\\
    Colon and rectum cancer  &  0  &  4  &  17\\
    \hline
  \end{tabular}
  \caption{Leave-one-out cross validatory criteria comparing predictive ability of three models $m$: a non-hierarchical model $m=1$ and two hierarchical models where the effects of gender and area either interact $(m=2)$ or are additive $(m=3)$ (or, for breast cancer, including only one gender).  Predictive ability is judged against the mortality data between the ages of 50 and 90.  The value of the criterion is presented as $LOOIC_m - LOOIC_1$, where $LOOIC_m = -2\sum_{i} elpd_{im}$, so that lower values are preferred, and values of 0 are shown for model $m=1$.  The non-hierarchical model is preferred in most cases.}
  \label{tab:cv}
\end{table}


\begin{thebibliography}{37}
\expandafter\ifx\csname natexlab\endcsname\relax\def\natexlab#1{#1}\fi
\expandafter\ifx\csname url\endcsname\relax
  \def\url#1{\texttt{#1}}\fi
\expandafter\ifx\csname urlprefix\endcsname\relax\def\urlprefix{URL: }\fi

\bibitem[{Barendregt et~al.(1998)Barendregt, {Van Oortmarssen}, Van~Hout, {Van
  Den Bosch} and Bonneux}]{barendregt:pmslt}
Barendregt, J.~J., {Van Oortmarssen}, G.~J., Van~Hout, B.~A., {Van Den Bosch},
  J.~M. and Bonneux, L. (1998) Coping with multiple morbidity in a life table.
\newblock \textit{Mathematical Population Studies}, \textbf{7}, 29--49.

\bibitem[{Barendregt et~al.(2003)Barendregt, Van~Oortmarssen, Vos and
  Murray}]{dismod2}
Barendregt, J.~J., Van~Oortmarssen, G.~J., Vos, T. and Murray, C. J.~L. (2003)
  {A generic model for the assessment of disease epidemiology: the
  computational basis of DisMod II}.
\newblock \textit{Population Health Metrics}, \textbf{1}, 4.

\bibitem[{Blakely et~al.(2020)Blakely, Moss, Collins, Mizdrak, Singh, Carvalho,
  Wilson, Geard and Flaxman}]{blakely2020proportional}
Blakely, T., Moss, R., Collins, J., Mizdrak, A., Singh, A., Carvalho, N.,
  Wilson, N., Geard, N. and Flaxman, A. (2020) Proportional multistate
  lifetable modelling of preventive interventions: concepts, code and worked
  examples.
\newblock \textit{International Journal of Epidemiology}, \textbf{49},
  1624--1636.

\bibitem[{Boshuizen et~al.(2017)Boshuizen, Nusselder, Plasmans, Hilderink,
  Snijders, Poos and {Van Gool}}]{boshuizen2017taking}
Boshuizen, H.~C., Nusselder, W.~J., Plasmans, M. H.~D., Hilderink, H.~H.,
  Snijders, B. E.~P., Poos, R. and {Van Gool}, C.~H. (2017) {Taking
  multi-morbidity into account when attributing DALYs to risk factors:
  comparing dynamic modeling with the GBD2010 calculation method}.
\newblock \textit{BMC Public Health}, \textbf{17}, 1--13.

\bibitem[{Briggs et~al.(2019)Briggs, Cobiac, Wolstenholme and
  Scarborough}]{briggs2019primetime}
Briggs, A. D.~M., Cobiac, L.~J., Wolstenholme, J. and Scarborough, P. (2019)
  {PRIMEtime CE}: a multistate life table model for estimating the
  cost-effectiveness of interventions affecting diet and physical activity.
\newblock \textit{BMC Health Services Research}, \textbf{19}, 1--19.

\bibitem[{Briggs et~al.(2016)Briggs, Wolstenholme, Blakely and
  Scarborough}]{briggs2016choosing}
Briggs, A. D.~M., Wolstenholme, J., Blakely, T. and Scarborough, P. (2016)
  Choosing an epidemiological model structure for the economic evaluation of
  non-communicable disease public health interventions.
\newblock \textit{Population Health Metrics}, \textbf{14}, 1--12.

\bibitem[{{British Heart Foundation}(2020)}]{bhf2020}
{British Heart Foundation} (2020) \textit{{Heart and Circulatory Disease
  Statistics 2020}}.
\newblock British Heart Foundation.

\bibitem[{Cecchini et~al.(2010)Cecchini, Sassi, Lauer, Lee, Guajardo-Barron and
  Chisholm}]{cecchini2010tackling}
Cecchini, M., Sassi, F., Lauer, J.~A., Lee, Y.~Y., Guajardo-Barron, V. and
  Chisholm, D. (2010) Tackling of unhealthy diets, physical inactivity, and
  obesity: health effects and cost-effectiveness.
\newblock \textit{The Lancet}, \textbf{376}, 1775--1784.

\bibitem[{Cox and Miller(1977)}]{cox:miller}
Cox, D.~R. and Miller, H.~D. (1977) \textit{{The Theory of Stochastic
  Processes}}, vol. 134.
\newblock CRC Press.

\bibitem[{{de S{\'a}} et~al.(2017){de S{\'a}}, Tainio, Goodman, Edwards,
  Haines, Gouveia, Monteiro and Woodcock}]{de2017health}
{de S{\'a}}, T.~H., Tainio, M., Goodman, A., Edwards, P., Haines, A., Gouveia,
  N., Monteiro, C. and Woodcock, J. (2017) {Health impact modelling of
  different travel patterns on physical activity, air pollution and road
  injuries for S{\~a}o Paulo, Brazil}.
\newblock \textit{Environment International}, \textbf{108}, 22--31.

\bibitem[{Flaxman(2019)}]{dismod:mr}
Flaxman, A.~D. (2019) {dismod-mr 1.1.1: Integrative Meta-Regression Framework
  for Descriptive Epidemiology.}
\newblock {Python package}.
\newblock \urlprefix\url{https://pypi.org/project/dismod-mr}.

\bibitem[{Flaxman et~al.(2015)Flaxman, Vos and Murray}]{flaxman2015}
Flaxman, A.~D., Vos, T. and Murray, C. J.~L. (2015) \textit{An integrative
  metaregression framework for descriptive epidemiology}.
\newblock University of Washington Press.

\bibitem[{{GBD 2019 Diseases and Injuries Collaborators}(2020)}]{gbd2019}
{GBD 2019 Diseases and Injuries Collaborators} (2020) {Global burden of 369
  diseases and injuries in 204 countries and territories, 1990--2019: a
  systematic analysis for the Global Burden of Disease Study 2019}.
\newblock \textit{The Lancet}, \textbf{396}, 1204--1222.

\bibitem[{Gelman et~al.(2013)Gelman, Carlin, Stern, Dunson, Vehtari and
  Rubin}]{gelman:bda3}
Gelman, A., Carlin, J.~B., Stern, H.~S., Dunson, D.~B., Vehtari, A. and Rubin,
  D. (2013) \textit{Bayesian Data Analysis}.
\newblock Chapman and Hall/CRC, 3rd edn.

\bibitem[{Hippisley-Cox et~al.(2017)Hippisley-Cox, Coupland and
  Brindle}]{qrisk3}
Hippisley-Cox, J., Coupland, C. and Brindle, P. (2017) {Development and
  validation of QRISK3 risk prediction algorithms to estimate future risk of
  cardiovascular disease: prospective cohort study}.
\newblock \textit{{BMJ}}, \textbf{357}.

\bibitem[{Iroz-Elardo et~al.(2020)Iroz-Elardo, Schoner, Fox, Brookes and
  Frank}]{iroz2020active}
Iroz-Elardo, N., Schoner, J., Fox, E.~H., Brookes, A. and Frank, L.~D. (2020)
  Active travel and social justice: Addressing disparities and promoting health
  equity through a novel approach to {Regional Transportation Planning}.
\newblock \textit{Social Science \& Medicine}, \textbf{261}, 113211.

\bibitem[{Jaller et~al.(2020)Jaller, Pourrahmani, Rodier, Maizlish and
  Zhang}]{jaller2020active}
Jaller, M., Pourrahmani, E., Rodier, C., Maizlish, N. and Zhang, M. (2020)
  {Active transportation and community health impacts of automated vehicle
  scenarios: an integration of the San Francisco Bay Area activity based travel
  demand model and the Integrated Transport and Health Impacts Model (ITHIM)}.
\newblock \textit{Cornell University CTECH Final Reports}.
\newblock \urlprefix\url{https://hdl.handle.net/1813/70173}.

\bibitem[{Keiding(1991)}]{keiding1991age}
Keiding, N. (1991) Age-specific incidence and prevalence: a statistical
  perspective.
\newblock \textit{Journal of the Royal Statistical Society: Series A
  (Statistics in Society)}, \textbf{154}, 371--396.

\bibitem[{Kypridemos et~al.(2016)Kypridemos, Allen, Hickey, Guzman-Castillo,
  Bandosz, Buchan, Capewell and O’Flaherty}]{kypridemos2016cardiovascular}
Kypridemos, C., Allen, K., Hickey, G.~L., Guzman-Castillo, M., Bandosz, P.,
  Buchan, I., Capewell, S. and O’Flaherty, M. (2016) Cardiovascular screening
  to reduce the burden from cardiovascular disease: microsimulation study to
  quantify policy options.
\newblock \textit{{BMJ}}, \textbf{353}.

\bibitem[{Lauer et~al.(2003)Lauer, R{\"o}hrich, Wirth, Charette, Gribble and
  Murray}]{lauer2003popmod}
Lauer, J.~A., R{\"o}hrich, K., Wirth, H., Charette, C., Gribble, S. and Murray,
  C. J.~L. (2003) {PopMod}: a longitudinal population model with two
  interacting disease states.
\newblock \textit{Cost Effectiveness and Resource Allocation}, \textbf{1},
  1--15.

\bibitem[{Mytton et~al.(2018)Mytton, Jackson, Steinacher, Goodman, Langenberg,
  Griffin, Wareham and Woodcock}]{mytton2018current}
Mytton, O.~T., Jackson, C., Steinacher, A., Goodman, A., Langenberg, C.,
  Griffin, S., Wareham, N. and Woodcock, J. (2018) {The current and potential
  health benefits of the National Health Service Health Check cardiovascular
  disease prevention programme in England: a microsimulation study}.
\newblock \textit{PLoS Medicine}, \textbf{15}, e1002517.

\bibitem[{Mytton et~al.(2017)Mytton, Tainio, Ogilvie, Panter, Cobiac and
  Woodcock}]{mytton2017modelled}
Mytton, O.~T., Tainio, M., Ogilvie, D., Panter, J., Cobiac, L. and Woodcock, J.
  (2017) The modelled impact of increases in physical activity: the effect of
  both increased survival and reduced incidence of disease.
\newblock \textit{European Journal of Epidemiology}, \textbf{32}, 235--250.

\bibitem[{{Office for National Statistics}(2019)}]{ons:cancer:survival}
{Office for National Statistics} (2019) {Cancer survival in England: national
  estimates for patients followed up to 2017}.
\newblock URL:
  \url{https://www.ons.gov.uk/peoplepopulationandcommunity/healthandsocialcare/conditionsanddiseases/bulletins/cancersurvivalinengland/nationalestimatesforpatientsfollowedupto2017}.

\bibitem[{O'Hagan et~al.(2006)O'Hagan, Buck, Daneshkhah, Eiser, Garthwaite,
  Jenkinson, Oakley and Rakow}]{ohagan:elic}
O'Hagan, A., Buck, C.~E., Daneshkhah, A., Eiser, J.~R., Garthwaite, P.~H.,
  Jenkinson, D.~J., Oakley, J.~E. and Rakow, T. (2006) \textit{{Uncertain
  Judgements: Eliciting Experts' Probabilities}}.
\newblock John Wiley \& Sons.

\bibitem[{Presanis et~al.(2013)Presanis, Ohlssen, Spiegelhalter and
  De~Angelis}]{presanis2013conflict}
Presanis, A.~M., Ohlssen, D., Spiegelhalter, D.~J. and De~Angelis, D. (2013)
  {Conflict diagnostics in directed acyclic graphs, with applications in
  Bayesian evidence synthesis}.
\newblock \textit{Statistical Science}, \textbf{28}, 376--397.

\bibitem[{Rehm et~al.(2009)Rehm, Mathers, Popova, Thavorncharoensap,
  Teerawattananon and Patra}]{rehm2009global}
Rehm, J., Mathers, C., Popova, S., Thavorncharoensap, M., Teerawattananon, Y.
  and Patra, J. (2009) Global burden of disease and injury and economic cost
  attributable to alcohol use and alcohol-use disorders.
\newblock \textit{The Lancet}, \textbf{373}, 2223--2233.

\bibitem[{Sax and Steiner(2013)}]{RJ-2013-028}
Sax, C. and Steiner, P. (2013) {Temporal disaggregation of time series}.
\newblock \textit{{The R Journal}}, \textbf{5}, 80--87.
\newblock \urlprefix\url{https://doi.org/10.32614/RJ-2013-028}.

\bibitem[{Scarborough et~al.(2016)Scarborough, Smolina, Mizdrak, Cobiac and
  Briggs}]{scarborough2016assessing}
Scarborough, P., Smolina, K., Mizdrak, A., Cobiac, L. and Briggs, A. (2016)
  {Assessing the external validity of model-based estimates of the incidence of
  heart attack in England: a modelling study}.
\newblock \textit{{BMC Public Health}}, \textbf{16}, 1--8.

\bibitem[{Scarborough et~al.(2011)Scarborough, Wickramasinghe, Bhatnagar and
  Rayner}]{bhftrends2011}
Scarborough, P., Wickramasinghe, K., Bhatnagar, P. and Rayner, M. (2011)
  \textit{Trends in coronary heart disease, 1961-2001}.
\newblock British Heart Foundation.

\bibitem[{Smolina et~al.(2012)Smolina, Wright, Rayner and
  Goldacre}]{smolina2012determinants}
Smolina, K., Wright, F.~L., Rayner, M. and Goldacre, M.~J. (2012) {Determinants
  of the decline in mortality from acute myocardial infarction in England
  between 2002 and 2010: linked national database study}.
\newblock \textit{{BMJ}}, \textbf{344}.

\bibitem[{{Stan Development Team}(2020)}]{rstan}
{Stan Development Team} (2020) {RStan}: the {R} interface to {Stan}.
\newblock R package version 2.21.2.
\newblock \urlprefix\url{http://mc-stan.org/}.

\bibitem[{Threlfall et~al.(2015)Threlfall, Meah, Fischer, Cookson, Rutter and
  Kelly}]{threlfall2015appraisal}
Threlfall, A.~G., Meah, S., Fischer, A.~J., Cookson, R., Rutter, H. and Kelly,
  M.~P. (2015) The appraisal of public health interventions: the use of theory.
\newblock \textit{Journal of Public Health}, \textbf{37}, 166--171.

\bibitem[{Vehtari et~al.(2020)Vehtari, Gabry, Magnusson, Yao, Bürkner,
  Paananen and Gelman}]{loor}
Vehtari, A., Gabry, J., Magnusson, M., Yao, Y., Bürkner, P.-C., Paananen, T.
  and Gelman, A. (2020) {loo: Efficient leave-one-out cross-validation and WAIC
  for Bayesian models}.
\newblock R package version 2.4.1.
\newblock \urlprefix\url{https://mc-stan.org/loo/}.

\bibitem[{Vehtari et~al.(2017)Vehtari, Gelman and Gabry}]{loocv}
Vehtari, A., Gelman, A. and Gabry, J. (2017) {Practical Bayesian model
  evaluation using leave-one-out cross-validation and WAIC}.
\newblock \textit{Statistics and Computing}, \textbf{27}, 1413--1432.

\bibitem[{Wood(2016)}]{jagam}
Wood, S.~N. (2016) {Just Another Gibbs Additive Modeller: interfacing JAGS and
  mgcv}.
\newblock \textit{Journal of Statistical Software}, \textbf{75}.

\bibitem[{Wood(2017)}]{wood2017generalized}
--- (2017) \textit{Generalized Additive Models: an Introduction with R}.
\newblock CRC, 2nd edn.

\bibitem[{Woodcock et~al.(2014)Woodcock, Tainio, Cheshire, O’Brien and
  Goodman}]{woodcock2014health}
Woodcock, J., Tainio, M., Cheshire, J., O’Brien, O. and Goodman, A. (2014)
  {Health effects of the London bicycle sharing system: health impact modelling
  study}.
\newblock \textit{{BMJ}}, \textbf{348}.

\end{thebibliography}
\end{document}